\documentclass[aps,prd,showpacs,notitlepage,nofootinbib,preprintnumbers,amsmath,amssymb]{revtex4-1}

\usepackage{subcaption}
\usepackage{graphics,graphicx}
\usepackage{dcolumn}
\usepackage{bm}
\usepackage{epsfig}
\usepackage[usenames]{color}
\usepackage{hyperref} 
\usepackage{mathbbol}
\usepackage{epstopdf}
\usepackage{simplewick}
\usepackage[utf8]{inputenc} 
\usepackage{slashed}
\usepackage{stackrel}
\usepackage{mathrsfs}
\usepackage{xcolor}
\usepackage{cancel}
\usepackage{mathtools}
\usepackage{multirow}
\usepackage{amsmath,amssymb,amsfonts,mathrsfs,bbold}
\usepackage[most]{tcolorbox}

\usepackage{physics}
\usepackage{subcaption}

\usepackage{ragged2e}
\DeclareCaptionJustification{justified}{\justifying}

\usepackage[normalem]{ulem}

\DeclareUnicodeCharacter{2009}{\,} 
\def\eq#1{{Eq.~(\ref{#1})}}
\def\fig#1{{Fig.~\ref{#1}}}
\def\sec#1{{Sec.~\ref{#1}}}
\def\tab#1{{Tab.~\ref{#1}}}
\newcommand{\ben}{\begin{eqnarray*}}
\newcommand{\een}{\end{eqnarray*}}
\newcommand{\un}[1]{\underline{#1}}

\newcommand{\bpsi}{\bar{\psi}}
\newcommand{\tord}{\textrm{T}}

\newcommand{\llangle}{\Big\langle \!\! \Big\langle}
\newcommand{\rrangle}{\Big\rangle \!\! \Big\rangle}

\newcommand{\as}{\alpha_s}
\newcommand{\kk}{\un{k}}

\newcommand{\xx}{\un{x}}

\makeatletter
\DeclareRobustCommand{\cev}[1]{%
  {\mathpalette\do@cev{#1}}%
}
\newcommand{\do@cev}[2]{%
  \vbox{\offinterlineskip
    \sbox\z@{$\m@th#1 x$}%
    \ialign{##\cr
      \hidewidth\reflectbox{$\m@th#1\vec{}\mkern4mu$}\hidewidth\cr
      \noalign{\kern-\ht\z@}
      $\m@th#1#2$\cr
    }%
  }%
}
\makeatother

\begin{document}

\title{Small-$x$ Asymptotics of the Leading-Twist Flavor Singlet Quark TMDs}

\author{Daniel Adamiak}
    \email[Email: ]{adamiak1@msu.edu}
        \affiliation{Jefferson Lab, Newport News, VA 23606, USA}
        \affiliation{Departmernt of Physics, Michigan State University, East Lansing, MI 48824, USA}
\author{M. Gabriel Santiago}
  \email[Email: ]{melvin.santiago@temple.edu}
	\affiliation{Department of Physics, Temple University, Philadelphia, PA 19122\\
    Department of Physics, Old Dominion University, Norfolk, VA 23606, USA \\
    Jefferson Lab, Newport News, Virginia 23606, USA \\
    Center for Nuclear Femtography, SURA, 1201 New York Avenue NW, Washington, DC 20005, USA}
\author{Yossathorn Tawabutr}
    \email[Email: ]{yossathorn.t@chula.ac.th}
    \affiliation{
Department of Physics, University of Jyv\"askyl\"a,  P.O. Box 35, 40014 University of Jyv\"askyl\"a, Finland
}
    \affiliation{
Helsinki Institute of Physics, P.O. Box 64, 00014 University of Helsinki, Finland
}
\affiliation{Department of Physics, Faculty of Science, Chulalongkorn University, 254 Phaya Thai Rd, Wang Mai, Pathum Wan, Bangkok 10330, Thailand}
	
\begin{abstract}
In this paper, we investigate the small-$x$ behavior of the flavor-singlet, leading-twist quark Transverse Momentum Dependent parton distribution functions (TMDs) using the Light-Cone Operator Treatment. This formalism allows us to express TMD operators at small $x$ in terms of polarized dipole amplitudes, enabling a systematic approach to their small-$x$ evolution. We derive the evolution equations for these TMDs and solve them within the large-$N_c$ approximation under the linearized, Double-Logarithmic Approximation (DLA), where $N_c$ represents the number of quark colors. Expanding on previous work on unpolarized and helicity TMDs, we present the small-$x$ asymptotics for a comprehensive set of TMDs, including the Sivers function, helicity worm-gear, transversity, pretzelosity, Boer-Mulders, and transversity worm-gear distributions. Our results provide a complete picture of the small-$x$ asymptotic behavior for all leading-twist flavor-singlet quark TMDs. We also discuss the implications of our findings for phenomenological applications and outline potential avenues for further research, particularly in understanding non-linear effects and extending beyond the DLA and large-$N_c$ approximations.
\end{abstract}

\maketitle


\section{Introduction}
\label{sec_int}
The intersection of spin in Quantum Chromodynamics (QCD) and small Bjorken-$x$ physics has seen significant development in recent years  \cite{Dominguez:2011wm, Dominguez:2011br, Kovchegov:2012ga,Kovchegov:2013cva,Zhou:2013gsa, Altinoluk:2014oxa, Boer:2015pni,Kovchegov:2015zha,Kovchegov:2015pbl, Dumitru:2015gaa, Szymanowski:2016mbq, Hatta:2016aoc, Hatta2016a,Hatta:2016khv,Boer:2016bfj,Balitsky:2016dgz, Kovchegov:2016zex, Kovchegov:2016weo, Kovchegov:2017jxc, Kovchegov:2017lsr, Dong:2018wsp, Benic:2018amn, Kovchegov:2018zeq, Kovchegov:2018znm, Chirilli:2018kkw, Altinoluk:2019wyu,Kovchegov:2019rrz, Boussarie:2019icw, Boussarie:2019vmk, Cougoulic:2019aja, Kovchegov:2020hgb, Cougoulic:2020tbc, Altinoluk:2020oyd, Kovchegov:2020kxg, Bacchetta:2020vty, Kovchegov:2021iyc, Chirilli:2021lif, Altinoluk:2021lvu, Adamiak:2021ppq, Kovchegov:2021lvz,Bondarenko:2021rbp,Banu:2021cla, Cougoulic:2022gbk,Kovchegov:2022kyy,Benic:2022qzv,Hatta:2022bxn,Borden:2023ugd, Li:2023tlw}. One main avenue in this work has been the inclusion of corrections beyond the usual saturation/Color Glass Condensate (CGC) framework \cite{Gribov:1984tu,Iancu:2003xm,Weigert:2005us,JalilianMarian:2005jf,Gelis:2010nm,Albacete:2014fwa,Kovchegov:2012mbw,Morreale:2021pnn,Wallon:2023asa}, in particular focusing on sub-eikonal and sub-sub-eikonal corrections \cite{Altinoluk:2014oxa,Kovchegov:2015pbl,Balitsky:2016dgz, Hatta:2016aoc, Kovchegov:2016zex, Kovchegov:2016weo, Kovchegov:2017jxc, Kovchegov:2017lsr, Kovchegov:2018znm, Kovchegov:2018zeq,Chirilli:2018kkw, Altinoluk:2019wyu, Kovchegov:2019rrz, Cougoulic:2019aja, Kovchegov:2020hgb, Cougoulic:2020tbc, Altinoluk:2020oyd, Chirilli:2021lif, Adamiak:2021ppq, Altinoluk:2021lvu, Kovchegov:2021lvz, Kovchegov:2021iyc, Cougoulic:2022gbk, Li:2023tlw} which allow for the exchange of spin quantum numbers in the CGC. This extension of the saturation framework allows one to go beyond the Collins-Soper-Sterman (CSS) evolution equations \cite{Collins:1981uw,Collins:1981uk,Collins:1981va,Collins:1984kg,Collins:1989gx}, which resum logarithms of the hard scale $Q^2$, and instead resum logarithms of $1/x$ to predict the small-$x$ behavior of TMDs. Using the Light Cone Operator Treatment (LCOT) originally developed in \cite{Kovchegov:2015pbl,Kovchegov:2017lsr, Kovchegov:2018znm, Kovchegov:2018zeq, Kovchegov:2021iyc}, the small-$x$ evolution equations of several leading-twist quark and gluon TMDs have been studied by constructing polarized dipole amplitudes containing spin-dependent sub-eikonal and sub-sub-eikonal corrections to the usual eikonal dipole amplitudes. Including the unpolarized quark TMD, which has small-$x$ behavior determined by the evolution equation for the Reggeon \cite{Kirschner:1983di,Kirschner:1985cb,Kirschner:1994vc,Kirschner:1994rq,Griffiths:1999dj,Itakura:2003jp}, all eight leading-twist quark TMDs have known small-$x$ asymptotics in the flavor non-singlet sector~\cite{Santiago:2023rfl}. This article expands on the work presented in~\cite{Santiago:2024iem}. In this work, we will study the flavor singlet sector of the TMDs, whose small-$x$ asymptotics were only known for the unpolarized and helicity TMDs. We will apply the LCOT to all six remaining TMD operator definitions in the massless quark regime, then find the asymptotics of the flavor singlet TMDs in the large-$N_c$ limit, further taking the linearized Double Logarithmic Approximation (DLA). Along with the physical insight gained from the finding of the asymptotic scaling of the leading-twist quark TMDs, we now have a complete set of equations to apply in phenomenological studies and global analyses, at least in the large-$N_c$, linearized DLA limit where $x$ is not yet so small that saturation effects become important.

\begin{table}[h] 
    \def\arraystretch{1.25}
    \centering
    \begin{tabular}{| c | c | c | c | c |} 
    \hline
    \multicolumn{5}{|c|}{Leading Twist Quark TMDs} \\ \hline
      & & \multicolumn{3}{|c|}{Quark Polarization} \\ \cline{3-5}
      & & U & L & T \\ \hline
      \multirow{3}{2.0cm}{Nucleon Polarization} & U & $f_1^{\textrm{S}} \sim x^{- \frac{4 \as N_c }{ \pi} \textrm{ln}(2)}$ &  & $h_1^{\perp \textrm{S}} \sim x$ \\ \cline{2-5}
      & L & & $g_1^{\textrm{S}} \sim x^{- 3.66 \sqrt{\as N_c / 2 \pi}}$ & $h_{1L}^{\perp \textrm{S}} \sim x$ \\ \cline{2-5}
      & T & $f_{1T}^{\perp \textrm{S}} \sim x^{-2.9 \sqrt{\as N_c / 4 \pi}}$ & $g_{1T}^{\textrm{S}} \sim x^{-2.9 \sqrt{\as N_c / 4 \pi}}$ & $h_1^{\textrm{S}} \sim h_{1T}^{\perp \textrm{S}} \sim  x^{1 - 2 \sqrt{\frac{\as N_c}{2\pi}}}$ \\ \cline{2-5}
      \hline
    \end{tabular} 
    \caption{The collected leading small-$x$ asymptotics for the leading-twist flavor singlet quark TMDs. The intercept of the unpolarized quark TMD $f_1$ is given by the solution for eikonal BFKL evolution \cite{Balitsky:1978ic} and the intercept for the helicity TMD $g_1$ was calculated using the LCOT in \cite{Cougoulic:2022gbk}. The remaining six intercepts are calculated using the LCOT in this work. }
    \label{tab_tmds}
\end{table}

The structure of the remainder of this paper is as follows: in \sec{sec:siv} we apply the LCOT to derive the small-$x$ expressions for the sub-eikonal contributions to the flavor singlet Sivers function $f_{1T}^{\perp}$ and the transverse worm-gear TMD $g_{1T}$. These expressions contain polarized dipole amplitudes which are very similar to those obtained in the flavor non-singlet case for the Sivers function \cite{Kovchegov:2022kyy} and in the flavor singlet case for the quark helicity TMD \cite{Cougoulic:2022gbk}. We obtain a set of closed evolution equations for these polarized dipole amplitudes in the large-$N_c$, linearized DLA and solve them numerically. Plugging the resulting asymptotics into the definition for the Sivers function and transversity worm-gear TMD yields
\begin{align}
    f_{1T}^{\perp \, \textrm{S}} (x \ll 1, k_T^2) \sim g_{1T}^{\textrm{S}} (x \ll 1, k_T^2) \sim \left(\frac{1}{x}\right)^{2.9 \sqrt{\frac{\as N_c}{2\pi}}} 
\end{align}

Next, in \sec{sec:tr_and_pr} we apply the LCOT to obtain the expressions for the sub-sub-eikonal contributions to the flavor singlet transversity and pretzelosity TMDs. The resulting polarized dipole amplitudes are very similar to the flavor non-singlet case for both TMDs \cite{Kovchegov:2018znm,Santiago:2023rfl}, but now the small-x evolution equations are more complicated. We analytically solve the equations in the large-$N_c$, linearized DLA and find that the new terms arising in the flavor singlet cases do not affect the asymptotics, so both TMDs actually match the asymptotics of their flavor non-singlet counterparts, 
\begin{align}
    h_{1T}^{\textrm{S}} (x \ll 1, k_T^2) \sim h_{1T}^{\textrm{NS}} (x \ll 1, k_T^2) \sim h_{1T}^{\perp \ \textrm{S}} (x \ll 1, k_T^2) \sim h_{1T}^{\textrm{NS}} (x \ll 1, k_T^2) \sim \left( \frac{1}{x} \right)^{-1 + 2 \sqrt{\frac{\as N_c}{2\pi}}} .
\end{align}
Finally, in \sec{sec:bmh} we apply the LCOT to obtain the expressions for the sub-sub-eikonal contributions to the flavor singlet Boer-Mulders $h_{1}^{\perp}$ and helicity worm-gear $h_{1L}$ TMDs. The polarized dipole amplitudes and corresponding evolution equations are again very similar to the flavor non-singlet case for both TMDs, so again we find the correspondence in that in the large-$N_c$, linearized DLA we have a match between the small-$x$ asymptotic scaling of the flavor singlet and flavor non-singlet TMDs as 
\begin{align}
    h_1^{\perp \, \textrm{S}} (x \ll 1, k_T^2) \sim h_1^{\perp \, \textrm{NS}} (x \ll 1, k_T^2) \sim h_{1L}^{\textrm{S}} (x \ll 1, k_T^2) \sim h_{1L}^{\textrm{NS}} (x \ll 1, k_T^2) \sim x ,
\end{align}
where we can see that these TMDs obey exactly their naive sub-sub-eikonal scaling as linear in $x$! We have collected all our new results along with the known results for the flavor singlet unpolarized quark TMD $f_1^{\textrm{S}}$ and the quark helicity TMD $h_1^{\textrm{S}}$ in Tab.~\ref{tab_tmds}. To conclude, in \sec{sec:conc} we summarize our findings and discuss the various interesting patterns and intricacies of the small-$x$ asymptotics of the flavor singlet, leading-twist quark TMDs which we have assembled here.

Throughout this paper we will make use of light-cone coordinates $u = (u^+ = x^0 + x^3, u^- = u^0 - u^3, \un{u})$, labelling the transverse part of a four-vector $u$ as $\un{u}$ except in the case of an integral measure, where it will be denoted as $u_{\perp}$, and in the case of the quark transverse momentum argument of a TMD, where we will use the conventional label $k_T$. We will also make use of Brodsky-Lepage (BL) spinors \cite{Lepage:1980fj}, specifically in the plus-minus reversed form defined as \cite{Kovchegov:2018znm,Kovchegov:2018zeq}
\begin{align}\label{anti BLspinors}
u_\sigma (p) = \frac{1}{\sqrt{p^-}} \, [p^- + m \, \gamma^0 +  \gamma^0 \, {\un \gamma} \cdot {\un p} ] \,  \rho (\sigma), \ \ \ v_\sigma (p) = \frac{1}{\sqrt{p^-}} \, [p^- - m \, \gamma^0 +  \gamma^0 \, {\un \gamma} \cdot {\un p} ] \,  \rho (-\sigma),
\end{align}
with $p^\mu = \left( \frac{{\un p}^2+ m^2}{p^-}, p^-, {\un p} \right)$ and
\begin{align}
  \rho (+1) \, = \, \frac{1}{\sqrt{2}} \, \left(
  \begin{array}{c}
      1 \\ 0 \\ -1 \\ 0
  \end{array}
\right), \ \ \ \rho (-1) \, = \, \frac{1}{\sqrt{2}} \, \left(
  \begin{array}{c}
        0 \\ 1 \\ 0 \\ 1
  \end{array}
\right) .
\end{align}
We will refer to these as anti BL spinors.

\section{ Sivers and worm-gear G TMDs}\label{sec:siv}

\subsection{Operator Treatment}

In this section we provide the steps in applying the Light Cone Operator Treatment (LCOT) to the Sivers function $f_{1T}^{\perp}$ and the helicity worm-gear TMD $g_{1T}$ and the definitions for the corresponding polarized dipole amplitudes.  The operator definition of the correlator containing the Sivers function is \cite{Meissner:2007rx}
\begin{align}\label{siv_1}
f_1^q (x,k_T^2) - \frac{\un{k} \cross \underline{S}_P}{M_P} f_{1 \: T}^{\perp \: q} (x,k_T^2) = \int \frac{\dd{r^-}\dd[2]{r_{\perp}}}{2  (2 \pi)^3} \, e^{i k \vdot r} \langle P, S_P | \bar{\psi}(0) \mathcal{U}[0,r] \frac{\gamma^+}{2} \psi(r) | P,S_P \rangle ,
\end{align}
where $f_1$ is the unpolarized quark density, $\ket{P,S_P}$ labels a hadron state of momentum $P$ and polarization $S_P$ in the transverse direction. The corresponding correlator for the worm-gear $g_{1T}$ is \cite{Meissner:2007rx}
\begin{align}
\frac{k_T \vdot S_P}{M_P} g_{1T} (x, k_T^2) = \int \frac{\dd{r}^- \dd[2]{r}_{\perp}}{2(2\pi)^3} \, e^{i k \vdot r} \bra{P, S_P} \bpsi (0) \mathcal{U} [0, r] \frac{\gamma^+ \gamma_5}{2} \psi(r)  \ket{P, S_P} .
\end{align}
In both correlators we take the future pointing Semi-Inclusive Deep Inelastic Scattering (SIDIS) staple gauge link $U [0,r] = V_{\un{0}} [0, \infty] V_{\un{r}} [\infty, r^-]$ with fundamental light-cone Wilson lines defined as
\begin{align}\label{Vline}
V_{\un{x}} [x^-_f,x^-_i] = \mathcal{P} \exp \left[ \frac{ig}{2} \int\limits_{x^-_i}^{x^-_f} \dd{x}^- A^+ (0^+, x^-, \un{x}) \right] .
\end{align}
We have dropped the transverse pieces of the gauge link out at infinity, so we must work in a non-singular gauge. Here we will take the proton `target' to be moving along the $x^+$ direction, and will work in $A^- = 0$ gauge for our calculations. We apply the LCOT \cite{Kovchegov:2018znm,Kovchegov:2018zeq}, rewriting the matrix elements as small-$x$ quasi-classical averages in the `target' proton state \cite{McLerran:1993ni,McLerran:1993ka,McLerran:1994vd,Kovchegov:1996ty,Balitsky:1997mk,Balitsky:1998ya,Kovchegov:2019rrz}
\begin{subequations}
\begin{align}
f_1^q (x,k_T^2) - \frac{\un{k} \cross \underline{S}_P}{M_P} f_{1 \: T}^{\perp \: q} (x,k_T^2) \subset \frac{2 p_1^+}{2 (2 \pi)^3} &\sum_X \int \dd{\xi^-} \dd[2]{\xi_{\perp}} \dd{\zeta^-} \dd[2]{\zeta_{\perp}} e^{i k \vdot (\zeta - \xi)}  \Big[\frac{\gamma^+}{2} \Big]_{\alpha \beta} \notag \\
&\times  \Big{\langle} \bar{\psi}_{\alpha} (\xi) V_{\underline{\xi}} [\xi^-,\infty] | X \rangle \langle X | V_{\underline{\zeta}} [\infty, \zeta^-] \psi_{\beta} (\zeta) \Big{\rangle} , \\
\frac{k_T \vdot S_P}{M_P} g_{1T} (x, k_T^2) \subset \frac{2 p_1^+}{2 (2 \pi)^3} \sum_X &\int \dd{\xi}^- \dd[2]{\xi}_{\perp} \dd{\zeta}^- \dd[2]{\zeta}_{\perp} e^{i k \vdot (\xi - \zeta)} \left[ \frac{\gamma^+ \gamma_5}{2} \right]_{\alpha \beta} \notag \\
&\times \langle \bpsi_{\alpha} (\xi) V_{\un{\xi}} [\xi^-, \infty] \ket{X} \bra{X} V_{\un{\zeta}} [\infty, \zeta^-] \psi_{\beta} (\zeta) \rangle .
\end{align}
\end{subequations}
Here we have inserted a sum over the complete set of states $\ket{X}$, and the outer angular brackets indicate the quasi-classical averaging over the proton wave function.

\begin{figure}[h]
\centering
\includegraphics[width=0.5\linewidth]{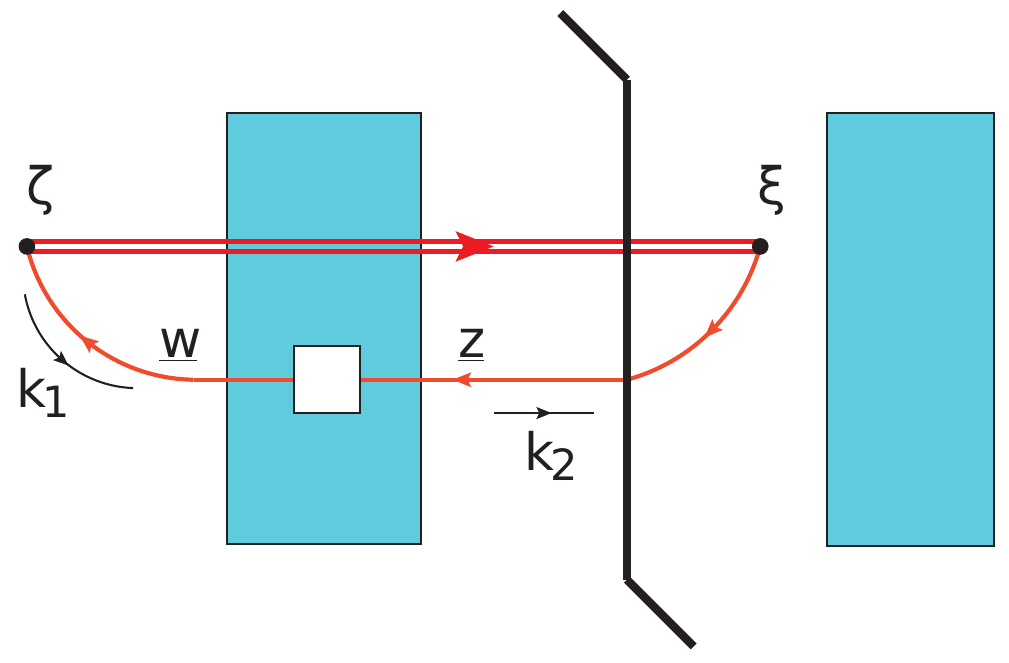}  
\caption{Example of the class of diagrams which give the leading sub-eikonal and sub-sub-eikonal corrections for the TMDs considered here. The anti-quark propagates from the position $\zeta$ to $\un{w}$ with momentum $k_1$, undergoes a sub-eikonal interaction with the proton which changes its transverse position from $\un{w}$ on the left of the shock wave to $\un{z}$ on the right of the shock wave. The anti-quark then propagates from $\un{z}$ to the position $\xi$ with momentum $k_2$. The shock wave is denoted by the blue (grey) rectangle, while the sub-(sub-) eikonal interaction with the shock wave is denoted by the white box. The double line represents the eikonal Wilson line encoding the interactions of the quark in the dipole.}
\label{FIG:diagbdet}
\end{figure}

\begin{figure}[h]
\centering
\includegraphics[width=1.0\linewidth]{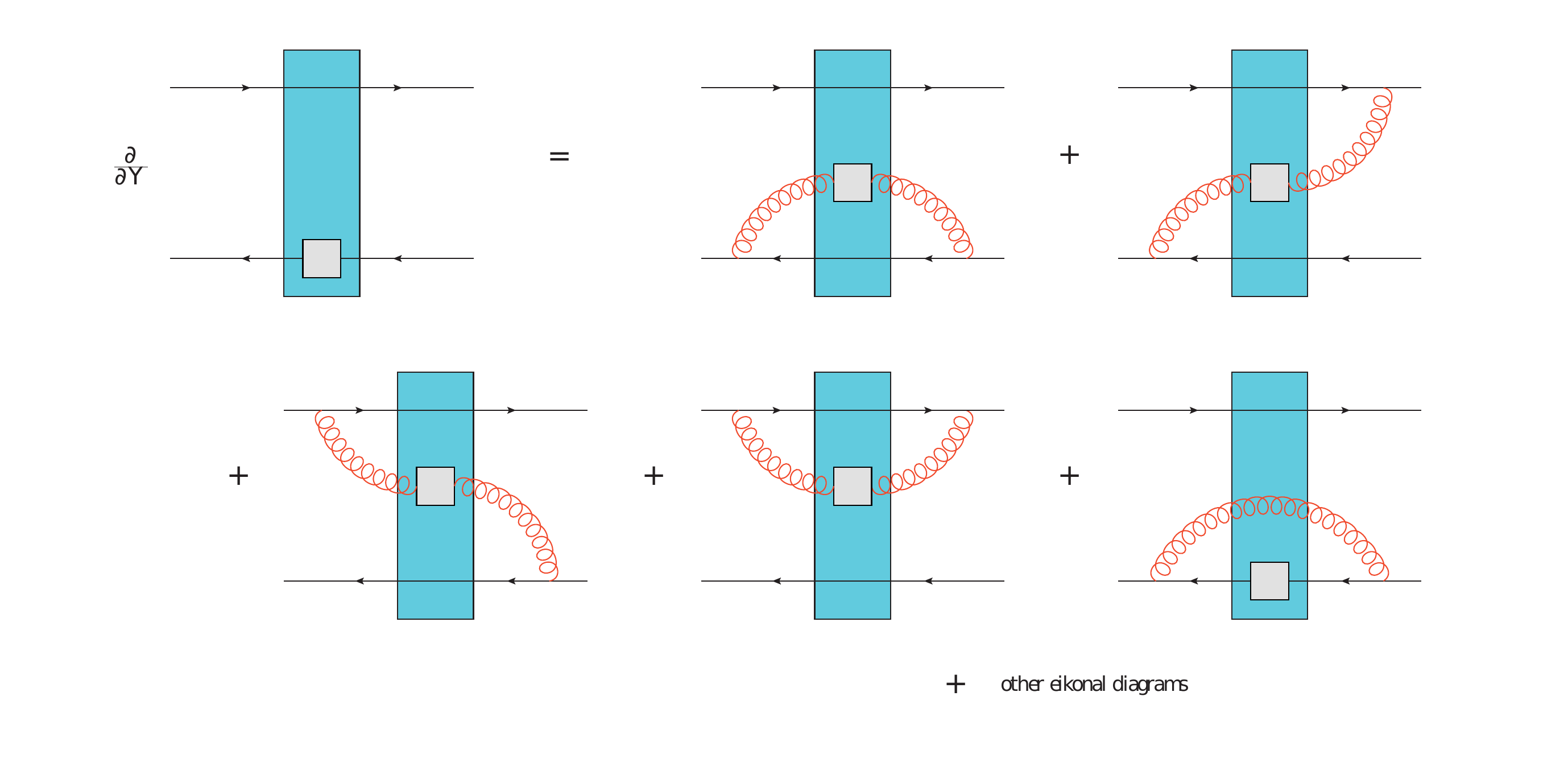}  
\caption{Typical classes of diagrams that appear in the evolution of polarized dipole amplitudes at large-$N_c$.}
\label{FIG:evo_diag}
\end{figure}

In the case of the Sivers function, the  leading small-$x$ eikonal contribution to the flavor non-singlet TMD is proportional to the spin dependent Odderon \cite{Boer:2015pni,Szymanowski:2016mbq,Dong:2018wsp}, but the flavor singlet TMD counterpart is zero \cite{Kovchegov:2022kyy}. In the case of $g_{1T}$ the eikonal contribution for both the flavor singlet and non-singlet TMDs vanish as the Odderon contribution cancels. Thus, we need to consider sub-eikonal (center of mass energy $\mathcal{O}(1/s)$ or equivalently Bjorken-x $\mathcal{O}(x_B)$ suppressed) corrections as the leading small-$x$ contribution to both the flavor singlet Sivers function and $g_{1T}$ that we consider here. One can show that the leading contribution to sub-eikonal order is given by the diagrams of the type shown in \fig{FIG:diagbdet} (\cite{Kovchegov:2018znm}), which yields 
\begin{align}
    &\frac{-2 p_1^+}{2(2 \pi)^3} \int \dd[2]{\zeta}_{\perp} \dd[2]{w}_{\perp} \dd[2]{z}_{\perp} \frac{\dd{k}_1^- \dd[2]{k}_{1 \perp}}{(2 \pi)^3}  \frac{e^{i \un{k}_1  \vdot (\un{w} - \un{\zeta}) + i \un{k} \vdot (\un{z} - \un{\zeta})} \theta(k_1^-)}{(x p_1^+ k_1^- + \un{k}_1^2) (x p_1^+ k_1^- + \un{k}^2)}  \\ 
    &\;\;\;\;\times \sum_{\sigma_1, \sigma_2} \bar{v}^{\alpha}_{\sigma_2} (k_2) \Gamma_{\alpha \beta} v^{\beta}_{\sigma_1} (k_1) \langle \tord \, V_{\un{\zeta}}^{ij} \bar{v}_{\sigma_1} (k_1) \hat{V}_{\un{z},\un{w}}^{\dagger \ ji} v_{\sigma_2} (k_2) \rangle \Big|_{k_1^- = k_2^-, k_1^2 = k_2^2 = 0, \un{k}_2 = -\un{k}} + (\text{c.c.}) .  \notag
\end{align}
where $\Gamma_{\alpha \beta}$ is the Dirac structure entering the TMD operator definition and $\alpha$ and $\beta$ are spinor indices. To proceed, one needs the following massless anti BL spinor products 
\begin{align}
    \bar{v}_{\sigma_2} (k_2) \frac{\gamma^+}{2} v_{\sigma_1}(k_1) = \frac{1}{ \sqrt{k_1^- k_2^-}} \big[ \delta_{\sigma_1,\sigma_2}  \, \underline{k}_1 \vdot \underline{k}_2 - i  \, \sigma_1 \delta_{\sigma_1, \sigma_2} \, \underline{k}_2 \cross \underline{k}_1 \big] .
\end{align}
Then, for the Sivers function we take $\Gamma = \gamma^+/2$ and make the replacement
\begin{align}
    \int d^2 z_\perp \, \bar{v}_{\sigma_1} (k_1) \left( \hat{V}_{{\un z}, {\un w}}^\dagger \right)^{ji}  v_{\sigma_2} (k_2)  \to 2 \sqrt{k_1^- \, k_2^-}  \,  \int d^2 z_\perp \, \left( V^{\textrm{pol} \, \dagger}_{{\un z}, {\un w}; \sigma_2 , \sigma_1} \right)^{ji} . 
\end{align}
Finally, we add the same correlator but for the anti-quark density to obtain the flavor singlet Sivers function correlator as
\begin{align}\label{siv_singlet_simp}
& - \frac{\un{k} \times \underline{S}_P}{M_P}   f_{1 \: T}^{\perp \: S} (x,k_T^2) \Big|_\textrm{sub-eikonal} 
=   \frac{16 N_c}{(2 \pi)^3} \int d^2 x_{10}   \, \frac{d^2 {k_{1 \perp}} }{(2\pi)^3}  \, \frac{e^{i (\un{k} + \underline{k}_1) \cdot \un{x}_{10} }}{\underline{k}_1^2 \,  \underline{k}^2}  \int\limits_\frac{\Lambda^2}{s}^1 \frac{dz}{z}  \\
&\;\;\;\;\;\;\;\;\times  \left\{ \underline{k}_1 \cdot \underline{k} \, (k -k_1)^i \, \left[ \epsilon^{ij} \, S_P^j \, x_{10}^2 \, F_A^S (x_{10}^2, z) + x_{10}^i \, {\un x}_{10} \times {\un S}_P \, F_B^S (x_{10}^2, z) +  \epsilon^{ij} \, x_{10}^j \, {\un x}_{10} \cdot {\un S}_P \, F_C^S (x_{10}^2, z) \right] \right. \notag \\ 
&\;\;\;\;+ \left.  i \, \underline{k}_1 \cdot \underline{k} \ {\un x}_{10} \times {\un S}_P \, F^{S \, [2]} (x_{10}^2, z) -  i \, \underline{k} \times \underline{k}_1 \, {\un x}_{10} \cdot {\un S}_P \, F^{S}_\textrm{mag} (x_{10}^2, z) \right\} , \notag
\end{align}
where the impact parameter integrated polarized dipole amplitudes 
\begin{subequations}\label{decomp}
\begin{align}
& \int d^2 b_\perp F^i_{10} = \epsilon^{ij} \, S_P^j \, x_{10}^2 \, F_A (x_{10}^2, z) + x_{10}^i \, {\un x}_{10} \times {\un S}_P \, F_B (x_{10}^2, z) +  \epsilon^{ij} \, x_{10}^j \, {\un x}_{10} \cdot {\un S}_P \, F_C (x_{10}^2, z), \label{decompA} \\
& \int d^2 b_\perp F^{[2]}_{10} = {\un x}_{10} \times {\un S}_P \, F^{[2]} (x_{10}^2, z), \label{decompF2} \\
& \int d^2 b_\perp F^\textrm{mag}_{10} = {\un x}_{10} \cdot {\un S}_P \, F_\textrm{mag} (x_{10}^2, z),
\end{align}
\end{subequations}
are defined in terms of the following polarized dipole amplitudes:
\begin{subequations}\label{siv_dipoles}
\begin{align}
& F^{S \, i}_{ {\un w}, {\un \zeta}} (z) = \frac{1}{2 N_c} \, \sum_f \, \mbox{Re} \, \llangle \tord  \tr \left[ V_{\underline{\zeta}} \, V^{i \, \dagger}_{{\un w}} \right] + \tord  \tr \left[ V^{i}_{{\un w}} \, V_{\underline{\zeta}}^\dagger \right]  \rrangle_1, \\
& F^{S \, [2]}_{ {\un w}, {\un \zeta}} (z) = \frac{1}{2 N_c} \, \sum_f \, \mbox{Im} \, \llangle \tord  \tr \left[ V_{\underline{\zeta}} \, V^{[2] \, \dagger}_{{\un w}; \un{k}, \un{k}_1} \right] - \tord  \tr \left[ V^{[2]}_{{\un w}; \un{k}, \un{k}_1} \, V_{\underline{\zeta}}^\dagger \right]  \rrangle_1 , \\ 
& F^{S \, \textrm{mag}}_{{\un w}, {\un \zeta}} (z) = \frac{1}{2 N_c} \, \sum_f \, \mbox{Re} \, \llangle \tord  \tr \left[ V_{\underline{\zeta}} \, V^{\textrm{mag} \, \dagger}_{{\un w}} \right] +  \tord  \tr \left[V^{\textrm{mag} }_{{\un w}} \, V_{\underline{\zeta}}^\dagger \right]  \rrangle_1 , 
\end{align}
\end{subequations}
which contain the polarized Wilson lines
\begin{subequations}\label{siv_wlines}
\begin{align}
V_{\un{x}}^i &= \frac{p_1^+}{4 \, s}  \int\limits_{-\infty}^{\infty} d{z}^- \ V_{\un{x}} [ \infty, z^-] \, \left[ \vec{D}^i_x -  \cev{D}_x^i \right]  \, V_{\un{x}} [ z^-, -\infty]\,, \label{ViV2a} \\
V_{\un{x}; \un{k}, \un{k}_1}^{[2]} &= \frac{i \, p_1^+}{8 \, s}  \int\limits_{-\infty}^{\infty} d{z}^- \ V_{\un{x}} [ \infty, z^-] \, \left[ (\vec{D}^i_x -  \cev{D}_x^i)^2 - (k_1^i -  k^i)^2 \right]  \, V_{\un{x}} [ z^-, -\infty]  \label{ViV2b} \\ 
&\;\;\;\;- \frac{g^2 \, p_1^+}{4 \, s} \, \int\limits_{-\infty}^{\infty} d{z}_1^- \int\limits_{z_1^-}^\infty d z_2^-  \ V_{\un{x}} [ \infty, z_2^-] \,  t^b \, \psi_{\beta} (z_2^-,\un{x}) \, U_{\un{x}}^{ba} [z_2^-,z_1^-] \, \left[ \frac{\gamma^+}{2}  \right]_{\alpha \beta} \bar{\psi}_\alpha (z_1^-,\un{x}) \, t^a \, V_{\un{x}} [ z_1^-, -\infty]\, , \notag \\
V_{\un{x}}^{\textrm{mag}} &= \frac{i \, g \, p_1^+}{2 \, s}  \int\limits_{-\infty}^{\infty} d{z}^- \ V_{\un{x}} [ \infty, z^-] \, F^{12} \, V_{\un{x}} [ z^-, -\infty] \\ 
&\;\;\;\;- \frac{g^2 \, p_1^+}{4 \, s} \, \int\limits_{-\infty}^{\infty} d{z}_1^- \int\limits_{z_1^-}^\infty d z_2^-  \ V_{\un{x}} [ \infty, z_2^-] \,  t^b \, \psi_{\beta} (z_2^-,\un{x}) \, U_{\un{x}}^{ba} [z_2^-,z_1^-] \, \left[ \frac{\gamma^+ \gamma^5}{2} \right]_{\alpha \beta} \bar{\psi}_\alpha (z_1^-,\un{x}) \, t^a \, V_{\un{x}} [ z_1^-, -\infty]\, . \notag 
\end{align}
\end{subequations}
 The double angle brackets $\llangle ... \rrangle_n$ scale out $n$ powers of energy. In particular, $\llangle ... \rrangle_n = (zs)^n\left\langle ... \right\rangle$. Hence, here we have $\llangle ... \rrangle_1$ scaling out one power of $zs$. For the $g_{1T}$ correlator we take $\Gamma = \gamma^+ \gamma_5 / 2$ to obtain

\begin{align}
\label{seg1t_fin}
\frac{\un{k}_x}{M}\,g_{1T}^{\text{S}}(x,k_{\perp})\Big|_{\textrm{sub-eik.}} &= \frac{N_c}{2\pi^4} \int \frac{dz}{z} \int d^2x_{10}  \int\frac{d^2k_1}{(2\pi)^2}   \,  e^{i(\kk+\kk_1)\cdot\xx_{10}} \, \frac{1}{k_{\perp}^2k_{1\perp}^2}       \\
&\;\;\;\;\;\times \Big[   (\kk_1\times\kk)(\kk-\kk_1)^i \, \Big[ \epsilon^{ij} \, S_P^j \, x_{10}^2 \, G_A^{\text{S}\, i}(x_{10}^2,z) + x_{10}^i \, {\un x}_{10} \times {\un S}_P \, G_B^{\text{S}\, i}(x_{10}^2,z) \notag \\
&+  \epsilon^{ij} \, x_{10}^j \, {\un x}_{10} \cdot {\un S}_P \, G_C^{\text{S}\, i}(x_{10}^2,z) \Big] - i(\kk_1\times\kk) \,  \ {\un x}_{10} \times {\un S}_P \, G^{\text{S}[2]}(x_{10}^2,z) +  i (\kk_1\cdot\kk) \,  {\un x}_{10} \cdot {\un S}_P \,  G^{\text{S}\,\text{mag}}(x_{10}^2,z) \Big]  ,         \notag  
\end{align}
where the impact parameter integrated polarized dipole amplitudes are defined analogously to \eqref{decomp}  in terms of the polarized dipole amplitudes defined as 
\begin{subequations}\label{siv_dipoles}
\begin{align}
& G^{S \, i}_{T \, {\un w}, {\un \zeta}} (z) = \frac{1}{2 N_c} \, \sum_f \, \mbox{Im} \, \llangle \tord  \tr \left[ V_{\underline{\zeta}} \, V^{i \, \dagger}_{{\un w}} \right] + \tord  \tr \left[ V^{i}_{{\un w}} \, V_{\underline{\zeta}}^\dagger \right]  \rrangle_1, \\
& G^{S \, [2]}_{T \, {\un w}, {\un \zeta}} (z) = \frac{1}{2 N_c} \, \sum_f \, \mbox{Re} \, \llangle \tord  \tr \left[ V_{\underline{\zeta}} \, V^{[2] \, \dagger}_{{\un w}; \un{k}, \un{k}_1} \right] - \tord  \tr \left[ V^{[2]}_{{\un w}; \un{k}, \un{k}_1} \, V_{\underline{\zeta}}^\dagger \right]  \rrangle_1 , \\ 
& G^{S \, \textrm{mag}}_{T \,{\un w}, {\un \zeta}} (z) = \frac{1}{2 N_c} \, \sum_f \, \mbox{Im} \, \llangle \tord  \tr \left[ V_{\underline{\zeta}} \, V^{\textrm{mag} \, \dagger}_{{\un w}} \right] +  \tord  \tr \left[V^{\textrm{mag} }_{{\un w}} \, V_{\underline{\zeta}}^\dagger \right]  \rrangle_1 . 
\end{align}
\end{subequations}
Here the polarized Wilson lines are the same as those entering the Sivers function above. Indeed, these are the same dipole amplitudes up to an exchange of ``Re'' and ``Im'' operators. Thus, up to differences in initial conditions, these dipole amplitudes will have identical operator-level evolution equations. Furthermore, the evolution kernels of these dipole amplitudes are known from studies of the quark helicity TMD and the flavor non-singlet Sivers function \cite{Cougoulic:2022gbk,Kovchegov:2022kyy}. They differ from the non-singlet evolution equations (Eq. (49) of \cite{Kovchegov:2022kyy}) in that terms in the first pair of integrals proportional to $x_{21}$ have the opposite sign. So we can immediately write down the large-$N_c$ evolution equations:
\begin{align}\label{F_i_ev}
& \epsilon^{ij} \, S_P^j \, x_{10}^2 \, F_A^{S} (x_{10}^2, z) + x_{10}^i \, {\un x}_{10} \times {\un S}_P \, F^{S}_B (x_{10}^2, z) +  \epsilon^{ij} \, x_{10}^j \, {\un x}_{10} \cdot {\un S}_P \, F^{S}_C (x_{10}^2, z)  \\ 
& = \epsilon^{ij} \, S_P^j \, x_{10}^2 \, F_A^{S\, (0)} (x_{10}^2, z) + x_{10}^i \, {\un x}_{10} \times {\un S}_P \, F^{S\, (0)}_B (x_{10}^2, z) +  \epsilon^{ij} \, x_{10}^j \, {\un x}_{10} \cdot {\un S}_P \, F^{S\, (0)}_C (x_{10}^2, z)  \notag \\ 
&\;\;\;\;+ \frac{\as N_c}{4 \pi^2}  \int\limits_{\frac{\Lambda^2}{s}}^z \frac{d z'}{z'} \int d^2 x_2 \, \Bigg\{ 2 \, \left[ \frac{\epsilon^{ij} x_{21}^j}{x_{21}^2} - \frac{\epsilon^{ij} x_{20}^j}{x_{20}^2} + 2 x_{21}^i \frac{{\un x}_{21} \times {\un x}_{20}}{x_{21}^2 \, x_{20}^2} \right] \left[  {\un x}_{21} \cdot {\un S}_P \, F^{S \, \textrm{mag}} (x_{21}^2, z')  +  {\un x}_{20} \cdot {\un S}_P \, F^{S \, \textrm{mag}} (x_{20}^2, z')   \right] \notag \\ 
&\;\;\;\;\;\;\;\;+ \left[ \delta^{ij} \left( \frac{3}{x_{21}^2} -  2 \, \frac{{\un x}_{20} \cdot {\un x}_{21}}{x_{20}^2 \, x_{21}^2} - \frac{1}{x_{20}^2} \right)  - 2 \frac{x_{21}^i \, x_{20}^j}{x_{21}^2 \, x_{20}^2} \left( 2 \frac{{\un x}_{20} \cdot {\un x}_{21}}{x_{20}^2} + 1 \right) + 2 \frac{x_{21}^i \, x_{21}^j}{x_{21}^2 \, x_{20}^2} \left( 2 \frac{{\un x}_{20} \cdot {\un x}_{21}}{x_{21}^2} + 1 \right) + 2 \frac{x_{20}^i \, x_{20}^j}{x_{20}^4} - 2 \frac{x_{21}^i \, x_{21}^j}{x_{21}^4}   \right] \notag \\
&\;\;\;\;\;\;\;\;\;\;\;\;\times \,  \left[  \epsilon^{jk} \, S_P^k \, x_{21}^2 \, F_A^{S} (x_{21}^2, z') + x_{21}^j \, {\un x}_{21} \times {\un S}_P \, F^{S}_B (x_{21}^2, z') +  \epsilon^{jk} \, x_{21}^k \, {\un x}_{21} \cdot {\un S}_P \, F^{S}_C (x_{21}^2, z')  \right. \notag \\ 
&\;\;\;\;\;\;\;\;\;\;\;\;\;\;\;\;+ \left.  \epsilon^{jk} \, S_P^k \, x_{20}^2 \, F_A^{S} (x_{20}^2, z') + x_{20}^j \, {\un x}_{20} \times {\un S}_P \, F^{S}_B (x_{20}^2, z') +  \epsilon^{jk} \, x_{20}^k \, {\un x}_{20} \cdot {\un S}_P \, F^{S}_C (x_{20}^2, z')  \right]  \Bigg\} \notag \\ 
&\;\;\;\;+ \frac{\as \, N_c}{2 \pi^2} \, \int\limits_{\frac{\Lambda^2}{s}}^z \frac{d z'}{z'} \, \int d^2 x_2 \, \frac{x_{10}^2}{x_{21}^2 \, x_{20}^2} \, \Bigg\{ \epsilon^{ij} \, S_P^j \, x_{21}^2 \, F_A^{S} (x_{21}^2, z') + x_{21}^i \, {\un x}_{21} \times {\un S}_P \, F^{S}_B (x_{21}^2, z') +  \epsilon^{ij} \, x_{21}^j \, {\un x}_{21} \cdot {\un S}_P \, F^{S}_C (x_{21}^2, z') \notag \\ 
&\;\;\;\;\;\;\;\;- \epsilon^{ij} \, S_P^j \, x_{10}^2 \, \Gamma_A^{S} (x_{10}^2, x_{21}^2, z') - x_{10}^i \, {\un x}_{10} \times {\un S}_P \, \Gamma^{S}_B (x_{10}^2, x_{21}^2, z') -  \epsilon^{ij} \, x_{10}^j \, {\un x}_{10} \cdot {\un S}_P \, \Gamma^{S}_C (x_{10}^2, x_{21}^2, z')  \Bigg\} \, .  \notag
\end{align}
Similarly, for $F^{\textrm{mag}}$, we have the result
\begin{align}\label{F_mag_ev}
F^{S \, \textrm{mag}}_{10} (z) &=  F^{S \, \textrm{mag} \, (0) }_{10} (z) + \frac{\as \, N_c}{2 \pi^2} \, \int\limits_{\frac{\Lambda^2}{s}}^z \frac{d z'}{z'} \, \int d^2 x_2 \, \Bigg\{ 2 \, \left[ \frac{1}{x_{21}^2} -  \frac{{\un x}_{21}}{x_{21}^2} \cdot \frac{{\un x}_{20}}{x_{20}^2} \right] \, \left[ F^{S \, \textrm{mag}}_{21} (z') +  F^{S \, \textrm{mag}}_{20} (z') \right]  \\ 
&\;\;\;\;\;\;\;\;+ \left[ 2 \frac{\epsilon^{ij} \, x_{21}^j}{x_{21}^4} - \frac{\epsilon^{ij} \, (x_{20}^j + x_{21}^j)}{x_{20}^2 \, x_{21}^2}  - \frac{2 \, {\un x}_{20} \times {\un x}_{21}}{x_{20}^2 \, x_{21}^2} \left( \frac{x_{21}^i}{x_{21}^2} - \frac{x_{20}^i}{x_{20}^2}\right) \right] \left[  F^{S \, i}_{21} (z') +  F^{S \, i}_{20} (z')   \right] \Bigg\}  \notag \\
&\;\;\;\;+ \frac{\as \, N_c}{2 \pi^2} \, \int\limits_{\frac{\Lambda^2}{s}}^z \frac{d z'}{z'} \, \int d^2 x_2 \, \frac{x_{10}^2}{x_{21}^2 \, x_{20}^2} \,  \Bigg\{ F^{S \, \textrm{mag}}_{12} (z')   - \Gamma^{S \, \textrm{mag}}_{10,21} (z')  \Bigg\} \, .  \notag
\end{align}
These equations arise from calculating emissions within the dipole amplitudes as depicted in \fig{FIG:evo_diag}. Specifically, one takes the contributions from all the depicted classes of diagrams except for those in class $\alpha$, as these contain polarized quark emissions which can be neglected at large-$N_c$. The calculation closely follows that in Sec. II C. of \cite{Kovchegov:2022kyy} for the sub-eikonal contribution to the flavor non-singlet Sivers function. These equations also contain `neighbor' dipole amplitudes $\Gamma^{S pol}$ which are analogous to those defined in \eq{siv_dipoles} but with a more complicated lifetime ordering, requiring another set of equations \cite{Kovchegov:2015pbl}. In particular, the neighbor dipole amplitude, $\Gamma^{S pol}_{10,21}(z')$, has lifetime $x^2_{21}z'$ despite its physical size being $x_{10}$. This is in contrast to an ordinary dipole amplitude, $F^{S pol}_{10}(z)$, whose lifetime, $x^2_{10}z$, is simply dictated by its physical size, $x_{10}$.

The DLA limit requires identifying the large logarithms that dominate the evolution equations \eqref{F_i_ev} and \eqref{F_mag_ev}. UV logarithms arise when we take the coincidence limits $2\to 1$ or $2\to 0$. IR logarithms arise when $x_{21}\gg x_{20},x_{10}$ or $x_{20}\gg x_{21},x_{10}$ . Looking at the first kernel in the sub-eikonal in \eqref{F_i_ev}, there are no UV logarithms associated with the $F^{S ~\textrm{mag}}$ terms. The IR logarithms are given by
\begin{align}\label{F_mag_IR_1}
    \frac{\alpha_s}{2\pi^2}(-2\pi x_{10}^i \underline{x}_{10}\times \underline{S}_p)\int\limits_{\frac{\Lambda^2}{s}}^z\frac{dz'}{z'}\int\frac{dx^2_{21}}{x_{21}^2} F^{S ~\textrm{mag}}(x_{21}^2) \, .
\end{align}
Moving on to the second kernel of the sub-eikonal term, the UV logarithms are the same as in the non-singlet case derived in \cite{Kovchegov:2022kyy} and will cancel against the UV logarithms in the eikonal term. The respective $x_{20}\gg x_{21},x_{10}$ and $x_{21}\gg x_{20},x_{10}$ IR limits of this kernel are given by
\begin{align}\label{F_Kernel_IR}
    &\int d^2x_{20}\left[ \delta^{ij}\frac{x_{10}^2}{x_{20}^4}(8\cos^2\theta_{20}-1)-8\frac{x_{20}^ix_{10}^jx_{10}}{x_{20}^5}\cos\theta_{20}+4\frac{x_{10}^ix_{10}^j}{x_{20}^4} \right],\\
    &\int d^2x_{21}\left[ \delta^{ij}\frac{x_{10}^2}{x_{10}^4}(3-8\cos^2\theta_{21})+8\frac{x_{21}^ix_{10}^jx_{10}}{x_{21}^5}\cos\theta_{21}-8\frac{x_{10}^ix_{21}^jx_{10}}{x_{21}^5}\cos\theta_{21}+2\frac{x_{10}^ix_{10}^j}{x_{20}^4} \right],
\end{align}
where $\theta_{ij}$ is the angle between the transverse vectors $\underline{x}_{i}$ and $\underline{x}_{j}$. When contracted with the kernels in \eqref{F_Kernel_IR}, the IR logarithms of the sub-eikonal term are given by
\begin{align}
    &\frac{\alpha_s}{4\pi^2} 2\pi  \left[ x_{10}^2\epsilon^{ij}S_P^j+x_{10}^i\underline{x}_{10}\times\underline{S}_P \right]\int\limits_{x_{10}^2}^{\frac{z}{z'}x_{10}^2}\frac{dx_{21}^2}{x_{21}^2}F_A^{S}(x_{12}^2,z')\,,\\
    &\frac{\alpha_s}{4\pi^2} \pi  \left[ x_{10}^2\epsilon^{ij}S_P^j-x_{10}^i\underline{x}_{10}\times\underline{S}_P \right]\int\limits_{x_{10}^2}^{\frac{z}{z'}x_{10}^2}\frac{dx_{21}^2}{x_{21}^2}F_B^{S}(x_{21}^2,z')\, ,\\
    &\frac{\alpha_s}{4\pi^2} \pi  \left[ x_{10}^2\epsilon^{ij}S_P^j+3x_{10}^i\underline{x}_{10}\times\underline{S}_P \right]\int\limits_{x_{10}^2}^{\frac{z}{z'}x_{10}^2}\frac{dx_{21}^2}{x_{21}^2}F_C^{S}(x_{21}^2,z')\, .
\end{align}
Moving on to identifying the large logarithms of \eqref{F_mag_ev}, the UV logarithms are the same as in the non-singlet case \cite{Kovchegov:2022kyy}. 
\begin{align}
    &\frac{\alpha_s}{2\pi^2} 2\pi \underline{x}_{10}\cdot\underline{S}_P \int\limits_{\frac{\Lambda^2}{s}}^z \frac{dz'}{z'}\int\limits_{\frac{1}{z's}}^{x_{10^2}}\frac{dx_{21}^2}{x_{21}^2}\Gamma^{S~\textrm{mag}}(x_{10}^2,x_{21}^2,z')\,,\\
     &\frac{\alpha_s}{2\pi^2} 2\pi \underline{x}_{10}\cdot\underline{S}_P \int\limits_{\frac{\Lambda^2}{s}}^z \frac{dz'}{z'}\int\limits_{\frac{1}{z's}}^{x_{10^2}}\frac{dx_{21}^2}{x_{21}^2}\Gamma^{S}_A(x_{10}^2,x_{21}^2,z')\,,\\
     &\frac{\alpha_s}{2\pi^2} (-\pi) \underline{x}_{10}\cdot\underline{S}_P \int\limits_{\frac{\Lambda^2}{s}}^z \frac{dz'}{z'}\int\limits_{\frac{1}{z's}}^{x_{10^2}}\frac{dx_{21}^2}{x_{21}^2}\Gamma^{S}_B(x_{10}^2,x_{21}^2,z')\,,\\
     &\frac{\alpha_s}{2\pi^2} 3\pi \underline{x}_{10}\cdot\underline{S}_P \int\limits_{\frac{\Lambda^2}{s}}^z \frac{dz'}{z'}\int\limits_{\frac{1}{z's}}^{x_{10^2}}\frac{dx_{21}^2}{x_{21}^2}\Gamma^{S}_C(x_{10}^2,x_{21}^2,z')\,,\\
    &\frac{\alpha_s}{2\pi^2} (-\pi) \underline{x}_{10}\cdot\underline{S}_P \int\limits_{\textrm{max}\left\lbrace\frac{\Lambda^2}{s},\frac{1}{sx_{10^2}}\right\rbrace}^z \frac{dz'}{z'}\int\limits_{\frac{1}{z's}}^{x_{10^2}}\frac{dx_{21}^2}{x_{21}^2}\Gamma^{S~\textrm{mag}}(x_{10}^2,x_{21}^2,z')\,.
\end{align}
The last term is the UV logarithm coming from the eikonal term in  \eqref{F_mag_ev}.

The IR logarithms vanished in the non-singlet case due to the Kernel being symmetric, but they will contribute here. They are given by
\begin{align}
    &\frac{\alpha_s}{2\pi^2} 2\pi \underline{x}_{10}\cdot\underline{S}_P \int\limits_{\frac{\Lambda^2}{s}}^z\frac{dz'}{z'}\int\limits_{x_{10}^2}^{\frac{z}{z'}x_{10}^2}\frac{dx_{21}^2}{x_{21}^2}F^{S\textrm{ mag}}(x_{21}^2,z')\,,\\
    &\frac{\alpha_s}{2\pi^2} 2\pi \underline{x}_{10}\cdot\underline{S}_P \int\limits_{\frac{\Lambda^2}{s}}^z\frac{dz'}{z'}\int\limits_{x_{10}^2}^{\frac{z}{z'}x_{10}^2}\frac{dx_{21}^2}{x_{21}^2}F^{S}_A(x_{21}^2,z')\,,\\
    &\frac{\alpha_s}{2\pi^2} (-\pi) \underline{x}_{10}\cdot\underline{S}_P \int\limits_{\frac{\Lambda^2}{s}}^z\frac{dz'}{z'}\int\limits_{x_{10}^2}^{\frac{z}{z'}x_{10}^2}\frac{dx_{21}^2}{x_{21}^2}F^{S}_B(x_{21}^2,z')\,,\\
    &\frac{\alpha_s}{2\pi^2} 3\pi \underline{x}_{10}\cdot\underline{S}_P \int\limits_{\frac{\Lambda^2}{s}}^z\frac{dz'}{z'}\int\limits_{x_{10}^2}^{\frac{z}{z'}x_{10}^2}\frac{dx_{21}^2}{x_{21}^2}F^{S}_C(x_{21}^2,z')\,.
\end{align}
Passing to the large-$N_c$ linearized approximation, we can write down the DLA evolution equations for the polarized dipoles.
\begin{subequations} \label{siv_g1t_dip_ev}
\begin{tcolorbox}[ams align]
    F_A^S(x_{10}^2,z) &= F_A^{S(0)}(x_{10}^2,z) + \frac{\alpha_s}{4\pi}\int\limits_{\frac{\Lambda^2}{s}}^{z}\frac{dz'}{z'}\int\limits_{\textrm{max}\left\lbrace x_{10}^2,\frac{1}{z's}\right\rbrace}^{\frac{z}{z'}x_{10}^2}\frac{dx_{21}^2}{x_{21}^2} \left[ 4F_A^S(x_{21}^2,z') +2F_B^S(x_{21}^2,z')+2F_C^S(x_{21}^2,z') \right]  \\
    F_B^S(x_{10}^2,z) &= F_B^{S(0)}(x_{10}^2,z) + \frac{\alpha_s}{4\pi}\int\limits_{\frac{\Lambda^2}{s}}^{z}\frac{dz'}{z'}\int\limits_{\textrm{max}\left\lbrace x_{10}^2,\frac{1}{z's}\right\rbrace}^{\frac{z}{z'}x_{10}^2}\frac{dx_{21}^2}{x_{21}^2}\\
    &\;\;\;\;\times \left[ 2F_A^S(x_{21}^2,z') -F_B^S(x_{21}^2,z')+3F_C^S(x_{21}^2,z')-2F^{S\textrm{ mag}}(x_{21}^2,z') \right]\nonumber\\
    F_C^S(x_{10}^2,z) &= F_C^{S(0)}(x_{10}^2,z)\\ 
    F^{S\textrm{ mag}}(x_{10}^2,z) &= F^{S \textrm{ mag}(0)}(x_{10}^2,z)  + \frac{\alpha_s}{4\pi}\int\limits_{\textrm{max}\left\lbrace\frac{\Lambda^2}{s},\frac{1}{sx_{10}^2}\right\rbrace}^{z}\frac{dz'}{z'}\int\limits_{\frac{1}{z's}}^{x_{10}^2}\frac{dx_{21}^2}{x_{21}^2}   \\
    &\;\;\;\;\times \left[2\Gamma^{S\textrm{ mag}}(x_{10}^2,x_{21}^2,z')+ 4\Gamma_A^S(x_{10}^2,x_{21}^2,z')-2\Gamma_A^S(x_{10}^2,x_{21}^2,z')+6\Gamma_A^S(x_{10}^2,x_{21}^2,z')\right] \nonumber \\
    + \frac{\alpha_s}{4\pi} &\int\limits_{\textrm{max}\left\lbrace\frac{\Lambda^2}{s},\frac{1}{z's}\right\rbrace}^{z}\frac{dz'}{z'}\int\limits_{\frac{1}{z's}}^{\frac{z}{z'}x_{10}^2}\frac{dx_{21}^2}{x_{21}^2} \left[ 2F^{S\textrm{ mag}}(x_{21}^2,z')+2F_A^S(x_{21}^2,z') -F_B^S(x_{21}^2,z')+3F_C^S(x_{21}^2,z')\right] ,\nonumber
\end{tcolorbox}
\end{subequations}
and the corresponding neighbor dipole amplitude equations
\begin{subequations} \label{siv_g1t_nei_ev}
\begin{tcolorbox}[ams align]
    \Gamma_A^S(x_{10}^2,x_{21}^2,z') &= F_A^{S(0)}(x_{10}^2,z') + \frac{\alpha_s}{4\pi}\int\limits_{\frac{\Lambda^2}{s}}^{z' \frac{x_{21}^2}{x_{10}^2}}\frac{dz''}{z''}\int\limits_{\textrm{max}\left\lbrace x_{10}^2,\frac{1}{z''s}\right\rbrace}^{\frac{z'}{z''}x_{21}^2}\frac{dx_{32}^2}{x_{32}^2} \\
    &\;\;\;\;\times \left[ 4F_A^S(x_{32}^2,z'') +2F_B^S(x_{32}^2,z'')+2F_C^S(x_{32}^2,z'') \right] \notag \\
    \Gamma_B^S(x_{10}^2,x_{21}^2,z') &= F_B^{S(0)}(x_{10}^2,z') + \frac{\alpha_s}{4\pi}\int\limits_{\frac{\Lambda^2}{s}}^{z' \frac{x_{21}^2}{x_{10}^2}}\frac{dz''}{z''}\int\limits_{\textrm{max}\left\lbrace x_{10}^2,\frac{1}{sz''}\right\rbrace}^{\frac{z'}{z''}x_{21}^2}\frac{dx_{32}^2}{x_{32}^2}\\
    &\;\;\;\;\times \left[ 2F_A^S(x_{32}^2,z'') -F_B^S(x_{32}^2,z'')+3F_C^S(x_{32}^2,z'')-2F^{S\textrm{ mag}}(x_{32}^2,z'') \right]\nonumber\\
    \Gamma_C^S(x_{10}^2, x_{21}^2,z') &= F_C^{S(0)}(x_{10}^2,z')\\ 
    \Gamma^{S\textrm{ mag}}(x_{10}^2, x_{21}^2,z') &= F^{S \textrm{ mag}(0)}(x_{10}^2,z')  + \frac{\alpha_s}{4\pi}\int\limits_{\textrm{max}\left\lbrace\frac{\Lambda^2}{s},\frac{1}{sx_{10}^2}\right\rbrace}^{z'\frac{x_{21}^2}{x_{10}^2}}\frac{dz''}{z''}\int\limits_{\frac{1}{z''s}}^{\textrm{min}\{x_{10}^2, x_{21}^2 \frac{z'}{z''}\}}\frac{dx_{32}^2}{x_{32}^2}  \\
    &\;\;\;\;\times \left[2\Gamma^{S\textrm{ mag}}(x_{10}^2,x_{32}^2,z'')+ 4\Gamma_A^S(x_{10}^2,x_{32}^2,z'')-2\Gamma_A^S(x_{10}^2,x_{32}^2,z'')+6\Gamma_A^S(x_{10}^2,x_{32}^2,z'')\right] \nonumber\\
    + \frac{\alpha_s}{4\pi} &\int\limits_{\textrm{max}\left\lbrace\frac{\Lambda^2}{s},\frac{1}{z''s}\right\rbrace}^{z' \frac{x_{21}^2}{x_{10}^2}}\frac{dz''}{z''}\int\limits_{\frac{1}{z''s}}^{\frac{z'}{z''}x_{21}^2}\frac{dx_{32}^2}{x_{232}^2} \left[ 2F^{S\textrm{ mag}}(x_{32}^2,z'')+2F_A^S(x_{32}^2,z'') -F_B^S(x_{32}^2,z'')+3F_C^S(x_{32}^2,z'')\right].\nonumber
\end{tcolorbox}
\end{subequations}
\subsection{Numerical Solution for the Sivers and worm-gear G TMD Evolution}
\begin{figure}
     \centering
     \begin{subfigure}[b]{0.32\textwidth}
         \centering
         \includegraphics[width=\textwidth]{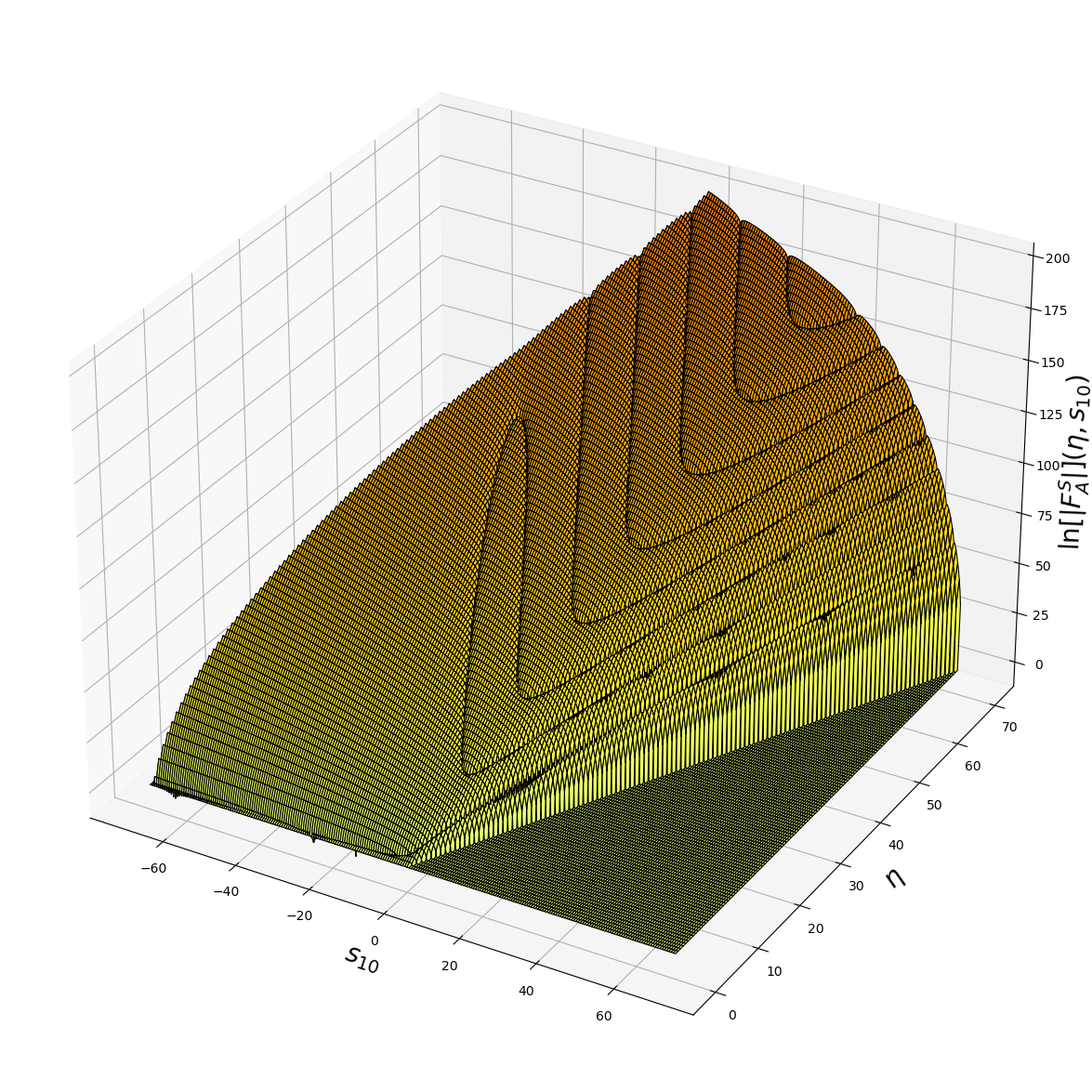}
         \caption{$\ln\left|F^S_A(s_{10},\eta)\right|$}
         \label{fig:QGG3dNf4_Q}
     \end{subfigure} 
  \;
     \begin{subfigure}[b]{0.32\textwidth}
         \centering
         \includegraphics[width=\textwidth]{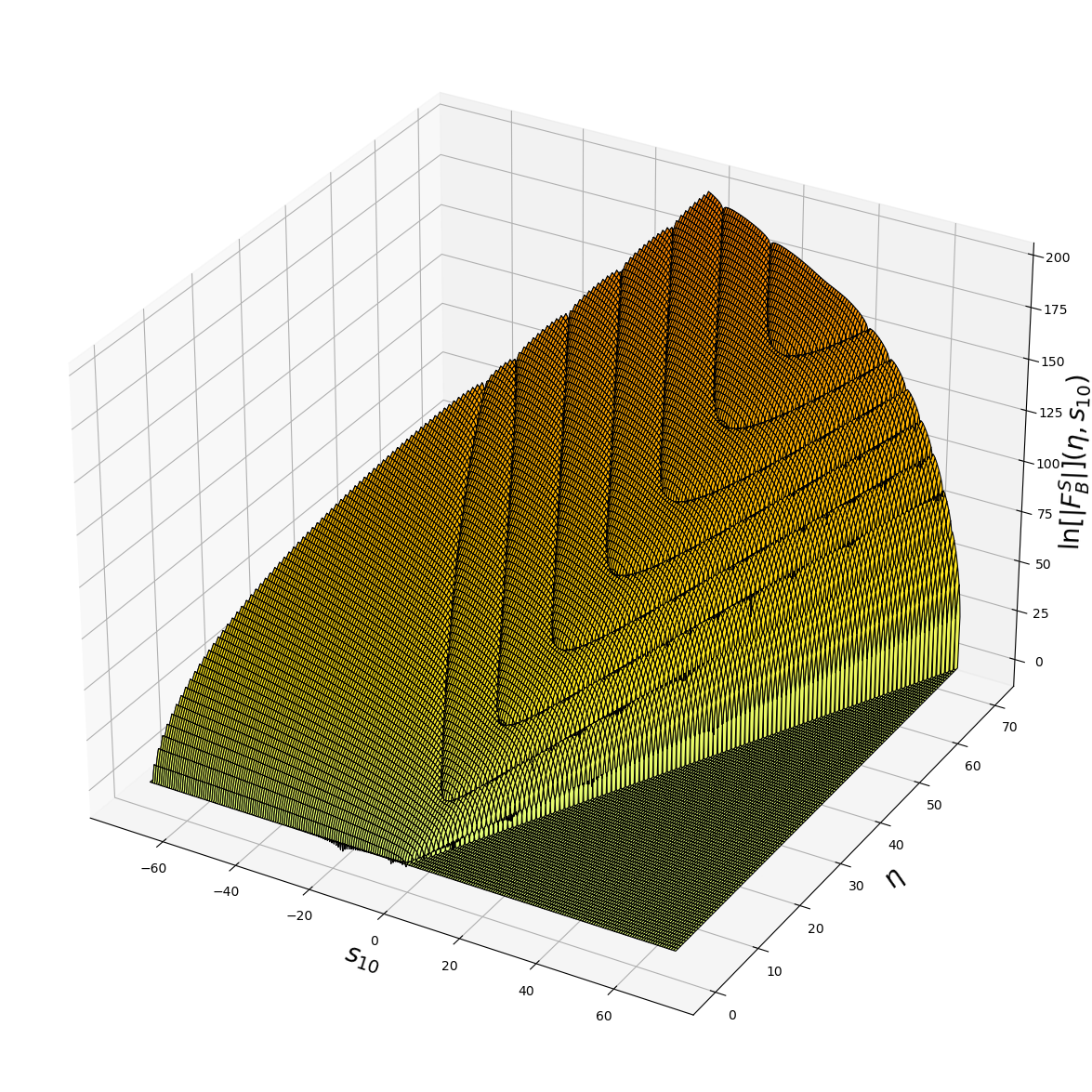}
         \caption{$\ln\left|F^S_B(s_{10},\eta)\right|$}
         \label{fig:QGG3dNf4_G2}
     \end{subfigure} 
 \;
     \begin{subfigure}[b]{0.32\textwidth}
         \centering
         \includegraphics[width=\textwidth]{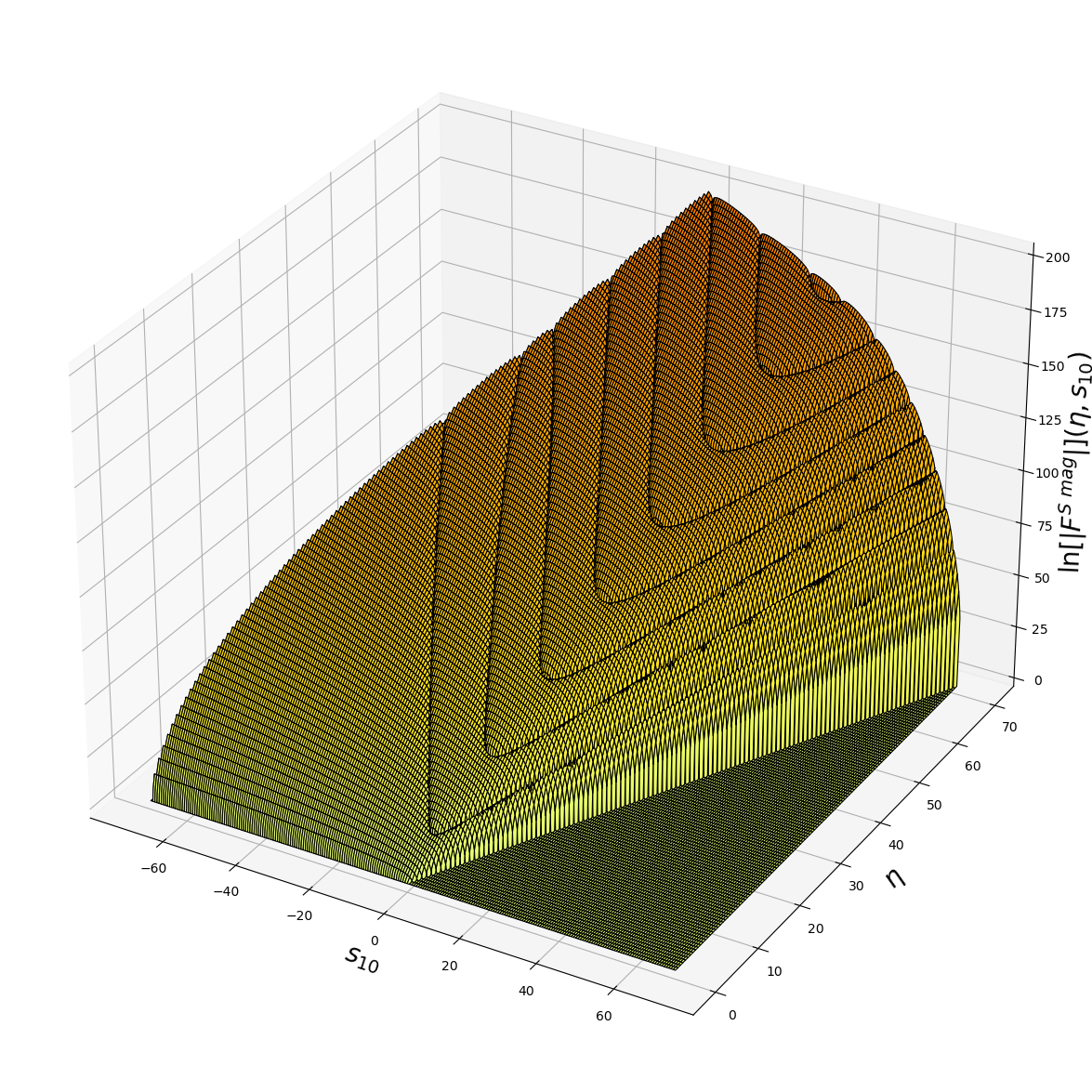}
         \caption{$\ln\left|F^{S~\rm{mag}}(s_{10},\eta)\right|$}
         \label{fig:QGG3dNf4_G}
     \end{subfigure}
     \caption{The plots of logarithms of the absolute values of polarized dipole amplitudes $F^S_A$, $F^S_B$ and $F^{S~\rm{mag}}$  versus $s_{10}$ and $\eta$, for the range $-\eta_{\max}\leq s_{10}\leq \eta_{\max}$ and $0\leq\eta\leq\eta_{\max}$ with $\eta_{\max}=70$. The amplitudes are computed numerically using step size $\Delta = 0.12$.}
     \label{fig:FABM3d}
\end{figure}

The evolution equations of the singlet Sivers and worm-gear G TMDs, \eqref{siv_g1t_dip_ev}, \eqref{siv_g1t_nei_ev}, are similar in structure to those of the non-singlet Sivers TMD \cite{Kovchegov:2022kyy}, so they permit the same numerical solution, with minor modifications, over the same grid. 
We make the changes of variables \cite{Kovchegov:2017jxc,Kovchegov:2020hgb,Cougoulic:2022gbk,Adamiak:2023okq,Adamiak:2023yhz}, 
\begin{align}\label{eta_s}
\eta^{(n)} &= \sqrt{\frac{\alpha_sN_c}{2\pi}}\,\ln\frac{z^{(n)}s}{\Lambda^2}\;\;\;\;\;\text{and}\;\;\;\;\;s_{ij} = \sqrt{\frac{\alpha_sN_c}{2\pi}}\,\ln\frac{1}{x^2_{ij}\Lambda^2}\,,
\end{align}
where the superscript, $(n)$, represents any number of primes, i.e. $\eta$, $\eta'$ and $\eta''$. We then discretize in these new variables:
\begin{align}\label{F_Gamma_disc}
    &F^{ S}_{ij} = F^{ S}(s_{10}=i\Delta, \eta=j\Delta)\;\;\;\;\;\text{and}\;\;\;\;\;\Gamma^{ S}_{ikj} = \Gamma^{ S}(s_{10}=i\Delta, s_{21}=k\Delta, \eta'=j\Delta) \, ,
\end{align}
where $\Delta$ is the step size in our $\eta^{(n)}$ and $s_{ij}$ grids. The step-size is kept the same for both sets of grids as it allows for faster numerical computations. The iterative solution to the evolution equations, \eqref{siv_g1t_dip_ev}, are then given by
\begin{subequations}
\begin{eqnarray}
    F^{S}_{A, ij} &= F_{A, i,j-1}^{S} + &\Delta^2\left[4F^S_{A,i,j-1}+2F^S_{B,i,j-1}+2F^S_{C,i,j-1} \right.\\
    &&+\left. \sum\limits_{j'=0}^{j-1}\left( 4F^S_{A,i+j'-j,j'}+2F^S_{B,i+j'-j,j'}+2F^S_{C,i+j'-j,j'}\right) \right]\nonumber\\
    F^{S}_{B, ij} &= F_{B, i,j-1}^{S} + &\Delta^2\left[2F^S_{A,i,j-1}-F^S_{B,i,j-1}+3F^S_{C,i,j-1}-2F^{S~\textrm{mag}}_{i,j-1} \right.\\
    &&+\left. \sum\limits_{j'=0}^{j-1}\left( 2F^S_{A,i+j'-j,j'}-F^S_{B,i+j'-j,j'}+3F^S_{C,i+j'-j,j'}-2F^{S~\textrm{mag}}_{i+j'-j,j'}\right) \right]\nonumber\\
    F^{S~\textrm{mag}}_{ij} &=  F^{S~\textrm{mag} (0)}_{i,j-1} +&\Delta^2\left[2F^S_{A,i,j-1}-F^S_{B,i,j-1}+3F^S_{C,i,j-1}+2F^{S~\textrm{mag}}_{i,j-1} \right.\\
    &&+\sum\limits_{j'=0}^{j-1} 2F^S_{A,i+j'-j,j'}-F^S_{B,i+j'-j,j'}+3F^S_{C,i+j'-j,j'}+2F^{S~\textrm{mag}}_{i+j'-j,j'}\nonumber\\
    &&+ \left.\sum\limits_{i'=i}^{j-1} 4\Gamma^S_{A,i,i',j-1}-2\Gamma^S_{B,i,i',j-1}+6\Gamma^S_{C,i,i',j-1}+2\Gamma^{S~\textrm{mag}}_{i,i',j-1}\right]\nonumber
\end{eqnarray}
\end{subequations}
Similarly, for the neighbor dipoles we have
\begin{subequations}
\begin{eqnarray}
    \Gamma^{S}_{A, ikj} &= \Gamma_{A, i,k,j-1}^{S} + &\Delta^2\left[4F^S_{A,i,\textrm{max}[0,i+j-k-1]}+2F^S_{B,i,\textrm{max}[0,i+j-k-1]}+2F^S_{C,i,\textrm{max}[0,i+j-k-1]} \right.\\
    &&+\left. \sum\limits_{j''=0}^{i+j-k-1}\left( 4F^S_{A,k+j''-j,j''}+2F^S_{B,k+j''-j,j''}+2F^S_{C,k+j''-j,j''}\right) \right]\nonumber\\
    \Gamma^{S}_{B, ikj} &= \Gamma_{B, i,k,j-1}^{S} + &\Delta^2\left[2F^S_{A,i,\textrm{max}[0,i+j-k-1]}-F^S_{B,i,\textrm{max}[0,i+j-k-1]}+3F^S_{C,i,\textrm{max}[0,i+j-k-1]}-2F^{S~\textrm{mag}}_{i,\textrm{max}[0,i+j-k-1]} \right.\nonumber\\
    &&+\left. \sum\limits_{j''=0}^{i+j-k-1}\left( 2F^S_{A,k+j''-j,j''}-F^S_{B,k+j''-j,j''}+3F^S_{C,k+j''-j,j''}-2F^{S~\textrm{mag}}_{k+j''-j,j''}\right) \right]\\
    \Gamma^{S~\textrm{mag}}_{ikj} &=  \Gamma^{S~\textrm{mag} (0)}_{i,k,j-1} +&\Delta^2\left[2F^S_{A,i,\textrm{max}[0,i+j-k-1]}-F^S_{B,i,\textrm{max}[0,i+j-k-1]}+3F^S_{C,i,\textrm{max}[0,i+j-k-1]}+2F^{S~\textrm{mag}}_{i,\textrm{max}[0,i+j-k-1]} \right.\nonumber\\
    &&+ \sum\limits_{j''=\textrm{max}[0,i+j-k-1]}^{j-1} 4\Gamma^S_{A,i,k+j''-j,j''}-2\Gamma^S_{B,i,k+j''-j,j''}+6\Gamma^S_{C,i,k+j''-j,j''}+2\Gamma^{S~\textrm{mag}}_{i,k+j''-j,j''}\nonumber\\
    &&+\sum\limits_{j''=0}^{i+j-k-1} 2F^S_{A,k+j''-j,j''}-F^S_{B,k+j''-j,j''}+3F^S_{C,k+j''-j,j''}+2F^{S~\textrm{mag}}_{k+j''-j,j''}\\
    &&+ \left.\sum\limits_{i''=k}^{j-1} 4\Gamma^S_{A,i,i'',j-1}-2\Gamma^S_{B,i,i'',j-1}+6\Gamma^S_{C,i,i'',j-1}+2\Gamma^{S~\textrm{mag}}_{i,i'',j-1}\right]\nonumber
\end{eqnarray}
\end{subequations}
These numerical evolution equations are iterated over $0\leq j\leq j_{max}$, $-j_{max}+j\leq i\leq j$, and $i\leq k<j$, where $\Delta j_{max} = \eta_{max}$, the largest value $\eta$ we can reasonably evolve to given our computational limitations. In previous works \cite{Kovchegov:2016zex} it was shown that the asymptotic behavior is largely independent of the initial condition, so we choose the simplest one: $F_A^{S(0)}=F_B^{S(0)}=F_C^{S(0)}=F^{S~\textrm{mag}(0)}=1$. In \fig{fig:FABM3d}, we plot the logarithm of the absolute value of the polarized dipole amplitudes after evolution for the constant initial condition.
What is surprising to see is the presence of oscillations, which appear as wedge-shaped ridges after taking the absolute value and logarithm. These were observed in the polarized dipole amplitudes contributing to the quark helicity TMD, \cite{Adamiak:2023okq}, but only in the large $N_c\& N_f$ limit for $N_f=6$. There is nothing in principle preventing these oscillations from appearing at large $N_c$; the balance of positive and negative terms on the right-hand side of \eqref{siv_g1t_dip_ev} cannot guarantee a consistent sign. The presence of oscillations makes extraction of the intercept trickier. In the absence of oscillations, the intercept could be extracted from the slope of the polarized dipole amplitudes in the $\eta$ direction at $s_{10}=0$ and large $\eta$. Because the oscillations confound this procedure, we need to extract the oscillating behavior in addition to the exponential growth so that we may isolate the different effects. In other words, at asymptotically small-$x$ or large $\eta$ (at $s_{10} = 0$), we model the polarized dipole amplitudes as 
\begin{eqnarray}\label{eq: oscillating ansatz}
    F^S(\eta,0) = e^{\alpha \eta}\cos{(\omega \eta + \phi)},
\end{eqnarray}
where we allow for different values of the intercept, $\alpha$, frequency, $\omega$, and phase, $\phi$, for each polarized dipole amplitude.
\\
\\
In order to extract the parameters in \eqref{eq: oscillating ansatz} and their associated uncertainties, we follow the procedure to handle the $N_f=6$ case employed in \cite{Adamiak:2023okq}. We perform the evolution on grids with $\eta_{\textrm{max}} \in \lbrace 10,20,30,40,50,60,70 \rbrace$ and the number of grid points, $N \in \lbrace 100,200,300,400,500,600 \rbrace$.  After extracting values for the three parameters for each of the dipole amplitudes, along with associated uncertainties at these various resolutions and values of $\eta_{\textrm{max}}$, we then proceed to take the continuum limit to determine the asymptotic values of these parameters. Taking the continuum limit is the procedure of fitting a polynomial to these parameter values as a function of the resolution, $\Delta = \frac{\eta_{\textrm{max}}}{N}$ and $\frac{1}{\eta_{\textrm{max}}}$. For example, 
\begin{eqnarray}
    \alpha(\Delta, 1/\eta_{\textrm{max}}) = \sum\limits_{i,j}c_{ij}\frac{\Delta^i}{\eta_{\textrm{max}}^j}
\end{eqnarray}
The continuum limit is the limit of $\Delta = 1/\eta_{\textrm{max}} \to 0$, i.e. the limit of infinite resolution and asymptotically large $\eta$. Using this choice of variables also conveniently gives the continuum value of the parameter as the constant term, $c_{00}$. One must be careful when deciding the degree of the polynomial in the continuum limit fit, as overfitting is a risk here. We analyze the coefficient of determination, $R^2$, choosing to stop at the polynomial that gives $R^2>1-10^{-5}$. We then combat overfitting further by pruning the parameters whose P-value for contributing is less than $5\%$. The results are summarized in table \ref{tab:params}

\begin{table}[]
    \centering
    \begin{tabular}{|c|c|c|c|}\hline
         Dipole Amplitudes & Intercept ($\alpha$)  & Frequency ($\omega$) & Phase ($\phi$)\\\hline
         $F^S_A$ & $2.821\pm 0.002$ & $0.211 \pm 0.003$ & $1.88 \pm 0.01$ \\\hline
         $F_B^S$ & $2.887 \pm 0.007$ & $0.197 \pm 0.003$ & $2.90 \pm 0.02$\\\hline
         $F^{S ~\rm{mag}}$ & $2.888 \pm 0.006$ & $0.179 \pm 0.002$ & $0.29 \pm 0.02$\\\hline
    \end{tabular}
    \caption{Continuum limit extraction of the parameters of the oscillating ansatz used to describe the polarized dipole amplitudes of Sivers and Worm-Gear G TMDs}
    \label{tab:params}
\end{table}
A word of caution is appropriate for interpreting the uncertainties of table \ref{tab:params}. The continuum limit depends on the model used to fit the numerical extractions and we have described one procedure for obtaining it. The uncertainties in the table are those extracted from Mathematica's non-linear model fit function, but when we adjust the model the intercept can change by about 0.1. 

Within one significant figure, we obtain the intercept of $\alpha = 3$. The singlet Sivers function grows slower than the non-singlet, where the intercept for the sub-eikonal contribution is 3.4. This is expected based on the naive time-reversal odd (T-odd) parity of the Sivers function, which leads to the charge conjugation odd (C-odd) contribution at eikonal order from the spin-dependent Odderon in the flavor non-singlet sector \cite{Boer:2015pni,Szymanowski:2016mbq,Dong:2018wsp} and a vanishing eikonal flavor singlet contribution. In contrast, the flavor non-singlet $g_{1T}$ is significantly suppressed compared to the flavor singlet sector studied here, being almost constant in $x$ and agreeing with usual expectations for small-$x$ physics. Indeed, the flavor singlet sector turns out to be dominant over the flavor non-singlet sector for all of the leading-twist T-even TMDs.

Beyond one significant figure, the intercept of $F^S_A$ disagrees with $F^S_B$ and $F^{S~\textrm{Mag}}$. One would expect that these values must eventually converge to the largest intercept, $\alpha = 2.89$, as $\eta\to\infty$. This is because the dipoles mix under evolution and should each be driven by the dipole that grows the fastest. The fact that this hasn't occurred yet, even when numerically computing up to $\eta_{\textrm{max}}=70$ suggests the presence of persistent pre-asymptotic behavior.

The return of the oscillations is a surprise and one can ask about the possibility of measuring them in an experiment. Much like the case of helicity, the answer is no. Not just because the period of oscillation is large in terms of rapidity of Bjorken-$x$, but because the presence of saturation will change the behavior of the dipoles before we can observe a full oscillation. At $Q^2=10 ~\textrm{GeV}^2$, we expect saturation to kick in below $x=10^{-5}$ \cite{Albacete_2005}. At $s_{10}=0$, this corresponds to $\eta>1.7$. Even the most quickly oscillating polarized dipole amplitude, $F_B^S$ has a period of $2\pi/0.186 = 33.8$. While the oscillations might have an effect on the pre-asymptotic behavior, we conclude that they don't practically affect the growth of the polarized dipole amplitudes.

The completion of a numeric solution to the evolution equations for the Sivers and Worm-Gear G singlet TMDs allows us to consider phenomenology in future work. To achieve flavor separation, one would need to rely on both the singlet and non-singlet evolution equations. Recall that our evolution is flavor blind - the flavor information enters in the initial condition, so the singlet evolution is applicable to the quark plus anti-quark flavor combination. 

An excellent observable candidate would be spin asymmetries in semi-inclusive deep inelastic scattering (SIDIS). We need to constrain the three light quarks and anti-quarks, so we need at least six independent observables. In SIDIS, at the relevant kinematic regime, we can measure two different outgoing hadrons, pions and kaons, that can be positively or negatively charged. We can also perform the scattering on both proton and neutron targets (neutron scattering is inferred from proton, deuteron, and helium scattering). That gives eight observables, enough to constrain the different flavors.

We need to carefully consider the kinematic range over which our observable is applicable. The hard scale is given by $Q^2$ and needs to be sufficiently large so that perturbation theory still works. In practice, $Q^2\sim 2~\rm{GeV}^2$ is sufficient. The $x_{Bj}$ cut is more subtle. The evolution of the Sivers and worm-gear G singlet TMDs is double logarithmic; the resummed quantity is $\alpha_s \ln^2 \frac{1}{x}$. Our formalism should be applicable when this resummed quantity $\mathcal{O}(1)$. This implies approximately that $x_{Bj}<10^{-1}$. This was the reasoning justifying the same cut in the study of small-$x$ helicity phenomenology \cite{Adamiak:2023yhz}. 

The same cut is not applicable for the non-singlet TMDs, per se, as this evolution is single logarithmic. This would imply the usual cut used in unpolarised small-x observables of $x_{Bj}<10^{-2}$. However, in the SIDIS observables, the singlet and non-singlet TMDs contribute simultaneously. In the region of $10^{-2}<x_{Bj}<10^{-1}$, the non-singlet might be small compared to large-$x$ effects, but the singlet contribution still should dominate over both. A consistent approach to implementing $x$-cuts would be to impose a cut on the data and evolution of the singlet at $x_{Bj} < 10^{-1}$, but a cut on the evolution of the non-singlet at $x_{Bj}<10^{-2}$. This prescription is not unique and we leave further discussion for the future.

The Sivers function has been measured in SIDIS by the HERMES, COMPASS and Jefferson Lab Hall A collaborations \cite{HERMES:2004mhh, COMPASS:2008isr, JeffersonLabHallA:2011vwy}. The relevant asymmetries for the worm-gear G TMD have also been measured in SIDIS by the same collaborations \cite{HERMES:2020ifk, COMPASS:2016led, Parsamyan:2018evv, JeffersonLabHallA:2011vwy}. There is enough data to perform a preliminary phenomenological study, but almost no data exists below $x_{Bj}=10^{-2}$, so the effects of singlet evolution will be hard to test and the non-singlet evolution won't be probed at all. The best prospect for measuring these asymmetries at small-$x$ is at the future Electron-Ion Collider, which should get measurements down to $x_{Bj} = 10^{-3}$ \cite{EICwhite}.

\section{Transversity and Pretzelocity TMDs}\label{sec:tr_and_pr}

Now we turn to the distributions of transversely polarized quarks within a transversely polarized hadron, the transversity $h_{1T}$ and pretzelosity $h_{1T}^{\perp}$ TMDs. The operator definition of the transversity TMD is \cite{Boussarie:2023izj}
\begin{align}\label{h1T_decomp}
    h_{1T} (x, k_T^2) + \frac{k_{\perp}^2}{M_P^2} h_{1T}^{\perp} (x, k_T^2) = \frac{1}{2} \sum_{S_p} S_p^j \int \frac{\dd{r}^- \dd[2]{r}_{\perp}}{2(2 \pi)^3} \, e^{ik\cdot r} \bra{P,S_P} \bar{\psi} (0) \mathcal{U} [0, r] \frac{i \sigma^{j+} \gamma_5}{2} \psi (r) \ket{P, S_P} ,
\end{align}
where $\sigma^{j+} = \frac{i}{2}[\gamma^j,\gamma^+] = -i\gamma^+\gamma^j$ is the usual Dirac matrices commutator. Without loss of generality we can take the proton spin $\un{S}_P$ vector to be along the $+\hat{x}$ direction to simplify Eq.~\eqref{h1T_decomp} to
\begin{align}
    h_{1T} (x, k_T^2) + \frac{k_{\perp}^2}{M_P^2} h_{1T}^{\perp} (x, k_T^2) = \int \frac{\dd{r}^- \dd[2]{r}_{\perp}}{2(2 \pi)^3} \, e^{ik\cdot r} \bra{P,S_P = +\hat{x}} \bar{\psi} (0) \mathcal{U} [0, r] \frac{ \gamma_5 \gamma^+ \gamma^1}{2} \psi (r) \ket{P, S_P = +\hat{x}}.
\end{align}
Applying the LCOT method, involving the propagation of an antiquark through the final state cut, to this operator yields the following expression in terms of flavor singlet polarized dipole amplitudes: \cite{Santiago:2023rfl}

\begin{align}
\label{trans_singlet_simp}
    &h_{1T}^{\textrm{S}} + \frac{k_{\perp}^2}{M_P^2} h_{1T}^{\perp \, \textrm{S}} (x, k_T^2) \Big{|}_{\textrm{sub-sub-eik}}  \subset\frac{ixN_c}{2\pi^4} \int \frac{dz}{z} \int d^2x_{10}  \int\frac{d^2k_1}{(2\pi)^2}  \,  e^{i(\kk+\kk_1)\cdot \xx_{10}} \, \frac{1}{k_{1\perp}^2k_{\perp}^2}\left(\frac{1}{k_{1\perp}^2} + \frac{1}{k_{\perp}^2}\right)          \\
    &\;\;\;\;\;\times \left\{ - \left[2(\underline{S}\cdot\kk_1)(\underline{S}\cdot\kk) - (\kk_1\cdot\kk)\right] H^{1 T, S}(x_{10}^2,z)  +  \left[(\underline{S}\cdot\kk_1)(\underline{S}\times\kk) + (\underline{S}\times\kk_1)(\underline{S}\cdot\kk)\right] H^{2 T, S}(x_{10}^2,z) \right\}    , \notag
\end{align}
Here, the sub-sub-eikonal dipole amplitudes are defined by the following correlators,
\begin{subequations}\label{trans_dipoles}
\begin{align}
& H^{1 T, S}_{10} (z) \equiv \frac{1}{2 N_c} \, \mbox{Im} \,  \llangle \tord \tr \left[ V_{\underline{0}} \, V^{\textrm{T} \, \dagger}_{{\un 1}} \right] - \tord \tr \left[ V_{\underline{0}}^\dagger \, V^{\textrm{T} }_{{\un 1}} \right] \rrangle_2, \\
& H^{2 T, S}_{10} (z) \equiv \frac{1}{2 N_c} \, \mbox{Re} \,  \llangle \tord \tr \left[ V_{\underline{0}} \, V^{\textrm{T} \, \perp \, \dagger}_{{\un 1}} \right] + \tord \tr \left[ V_{\underline{0}}^\dagger \, V^{\textrm{T} \, \perp}_{{\un 1}} \right]  \rrangle_2 , 
\end{align}
\end{subequations}
with the impact parameter integrated dipole amplitudes entering \eq{trans_singlet_simp} given by
\begin{subequations}\label{trans_b_int}
\begin{align}
    &\int \dd[2]{b}_{\perp} H^{1 T, S}_{10} (z) = H^{1 T, S} (x_{10}^2,z) \\
    &\int \dd[2]{b}_{\perp} H^{2 T, S}_{10} (z) = H^{2 T, S} (x_{10}^2,z) ,
\end{align}
\end{subequations}
for $\underline{b}=(\underline{x}_0+\underline{x}_1)/2$. Throughout these definitions, we assumed that the impact parameter integrated dipole amplitudes contain no integer powers of $x_{10}$. The polarized Wilson lines entering these polarized dipole amplitudes are
\begin{subequations}\label{t_wlines}
\begin{align}
V_{\un{x}}^\textrm{T} \equiv & \, \frac{g^2 \, (p_1^+)^2}{16 \, s^2} \, \int\limits_{-\infty}^{\infty} d{z}_1^- \int\limits_{z_1^-}^\infty d z_2^-  \ V_{\un{x}} [ \infty, z_2^-] \, t^b \, \psi_{\beta} (z_2^-,\un{x}) \, U_{\un{x}}^{ba} [z_2^-,z_1^-] \, \Bigg[ \left[ i \gamma^5 \underline{S} \cdot \cev{\underline{D}}_{x} - \underline{S} \times \cev{\underline{D}}_{x} \right] \, \gamma^+ \gamma^- \\ 
& +  \left[ i \gamma^5 \underline{S} \cdot \underline{D}_{x}  - \underline{S} \times \underline{D}_{x} \right] \gamma^- \gamma^+ \Bigg]_{\alpha \beta} \bar{\psi}_\alpha (z_1^-,\un{x}) \, t^a \, V_{\un{x}} [ z_1^-, -\infty]  , \notag \\
V_{\un{x}}^{\textrm{T} \, \perp} \equiv & \, - \frac{g^2 \, (p_1^+)^2}{16 \, s^2} \, \int\limits_{-\infty}^{\infty} d{z}_1^- \int\limits_{z_1^-}^\infty d z_2^-  \ V_{\un{x}} [ \infty, z_2^-] \, t^b \, \psi_{\beta} (z_2^-,\un{x}) \, U_{\un{x}}^{ba} [z_2^-,z_1^-] \,  \Bigg[ \left[ i \underline{S} \cdot \cev{\underline{D}}_{x} - \gamma^5 \underline{S} \times \cev{\underline{D}}_{x} \right]  \, \gamma^+ \gamma^-  \\ 
& +  \left[ i \underline{S} \cdot \underline{D}_{x}  - \gamma^5 \underline{S} \times \underline{D}_{x} \right] \, \gamma^- \gamma^+ \Bigg]_{\alpha \beta}   \bar{\psi}_\alpha (z_1^-,\un{z}_1) \, t^a \, V_{\un{x}} [ z_1^-, -\infty]  . \notag
\end{align}
\end{subequations}

Similarly, the operator definition of the pretzelosity TMD is
\begin{align}
    \frac{(k_T \vdot S_P) \, (k_T \cross S_P)}{M_P^2} h_{1T}^{\perp \, q} (x, k_T^2) = \epsilon^{ij} \, \frac{1}{2}\sum_{S_P} S_P^i  \int \frac{\dd{r}^- \dd[2]{r}_{\perp}}{2(2 \pi)^3} \bra{P,S_P} \bar{\psi} (0) \mathcal{U} [0, r] \frac{i \sigma^{j+} \gamma_5}{2} \psi (r) \ket{P, S_P},
\end{align}
for which we can explicitly take the proton spin along the $y$-direction, $S_P = +\hat{y}$, to simplify to
\begin{align} \label{pretz_def2}
    -\frac{k_T^x k_T^y}{M_P^2} h_{1T}^{\perp \, q} (x, k_T^2) = \int \frac{\dd{r}^- \dd[2]{r}_{\perp}}{2(2 \pi)^3} \bra{P,S_P = +\hat{y}} \bar{\psi} (0) \mathcal{U} [0, r] \frac{\gamma_5 \gamma^+ \gamma^1 }{2} \psi (r) \ket{P, S_P = +\hat{y}}.
\end{align}
This yields the flavor singlet expression

\begin{align}
\label{pretz_singlet_simp}
    & -\frac{k_T^x k_T^y}{M_P^2} h_{1T}^{\perp \, \textrm{S}} (x, k_T^2) \Big{|}_{\textrm{sub-sub-eik}}  \subset  \frac{ixN_c}{2\pi^4} \int \frac{dz}{z} \int d^2x_0 \,  d^2x_1  \int\frac{d^2k_1}{(2\pi)^2}  \,  e^{i(\kk+\kk_1)\cdot \xx_{10}} \, \frac{1}{k_{1\perp}^2k_{\perp}^2}\left(\frac{1}{k_{1\perp}^2} + \frac{1}{k_{\perp}^2}\right)     \notag     \\
    &\;\;\;\;\;\times \left\{  \left[2(\underline{S}\cdot\kk_1)(\underline{S}\cdot\kk) - (\kk_1\cdot\kk)\right] H^{1 \perp, S}_{10}(x_{10}^2, z)  -  \left[(\underline{S}\times\kk_1)(\underline{S}\cdot\kk) + (\underline{S}\cdot\kk_1)(\underline{S}\times\kk)\right] H^{2 \perp, S}_{10}(x_{10}^2, z) \right\}
\end{align}
with polarized dipole amplitudes 
\begin{subequations}\label{pretz_dipoles}
\begin{align}
& H^{1 \perp, S}_{10}(z) \equiv \frac{1}{2 N_c} \, \mbox{Im} \,  \llangle \tord \tr \left[ V_{\underline{0}} \, V^{\textrm{T} \, \dagger}_{{\un 1}} \right] - \tord \tr \left[ V_{\underline{0}}^\dagger \, V^{\textrm{T} }_{{\un 1}} \right] \rrangle_2, \\
& H^{2 \perp, S}_{10}(z) \equiv \frac{1}{2 N_c} \, \mbox{Re} \,  \llangle \tord \tr \left[ V_{\underline{0}} \, V^{\textrm{T} \, \perp \, \dagger}_{{\un 1}} \right] + \tord \tr \left[ V_{\underline{0}}^\dagger \, V^{\textrm{T} \, \perp}_{{\un 1}} \right]  \rrangle_2 , 
\end{align}
\end{subequations}
which are integrated over impact parameter as
\begin{subequations}\label{eqn:54}
\begin{align}
    &\int \dd[2]{b}_{\perp} H^{1 \perp, S}_{10} (z) = H^{1 \perp, S} (x_{10}^2, z)  \\
    &\int \dd[2]{b}_{\perp} H^{2 \perp, S}_{10} (z) = H^{2 \perp, S} (x_{10}^2, z) \,  .
\end{align}
\end{subequations}

\begin{figure}[h]
\centering
\includegraphics[width=0.5\linewidth]{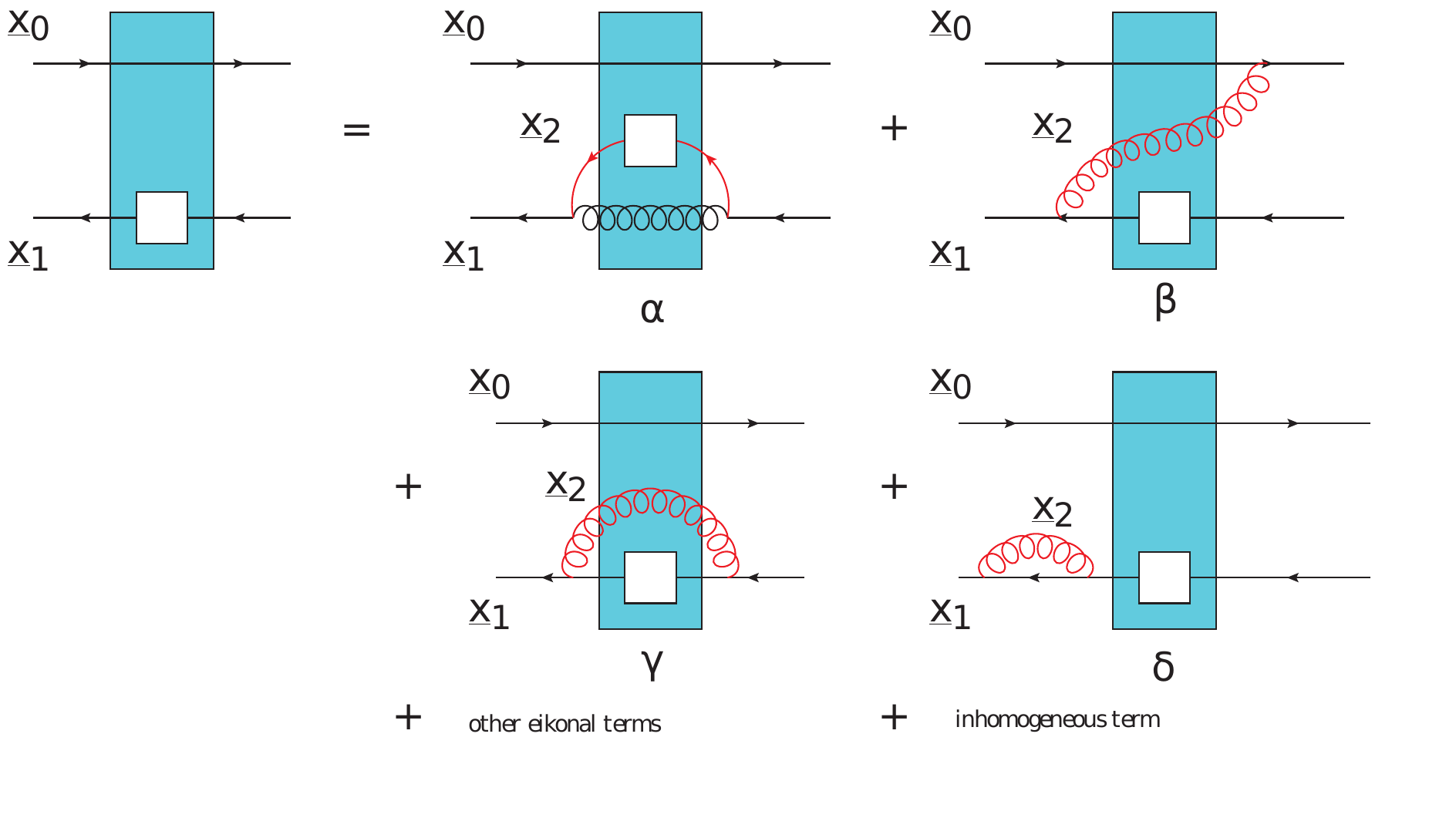}  
\caption{Classes of diagrams contributing to the evolution of polarized dipole amplitudes where only polarized quark emission (class $\alpha$) and eikonal gluon emissions (remaining classes) are allowed.}
\label{FIG:H_evolution}
\end{figure}

As these two TMDs contain the same polarized Wilson lines, $V_{\un{x}}^{\textrm{T}}$ and $V_{\un{x}}^{\textrm{T} \, \perp}$, inside nearly identical color trace structures that make up the respective polarized dipole amplitudes, the small-$x$ evolution equations at the operator level will be the same. We can neglect polarized gluon exchange contributions to the Wilson lines as these come with a factor of the light quark mass at the sub-sub-eikonal level and thus cannot generate DLA evolution. This means we can also neglect diagrams involving polarized gluon emissions in the derivation of the evolution equations, so we only have to consider the polarized quark emission diagrams and the eikonal, unpolarized gluon emission diagrams as shown in \fig{FIG:H_evolution}. The operator evolution equations have the same form for all four dipole amplitudes, so we can look at a single equation, noting that we will pick up a second equation for each dipole if we have neighbor dipole contributions from UV divergences. The equation has the form
\begin{align}
    H_{10}^{1 T, S} (z) &= H^{1 T, S \, (0)}_{10}(z) + \frac{\alpha_s N_c}{2\pi^2} \int\limits_{\frac{\Lambda^2}{s}}^z \frac{\dd{z}'}{z'} \int \dd[2]{x}_2 \frac{x_{10}^2}{x_{21}^2 x_{20}^2}  \, \mbox{Re} \, \llangle \frac{1}{N_c^2} \,  \tord \tr \left[ V_{\un{0}} t^a V_{\un{1}}^{\textrm{T} \dagger} t^b \right] \, \left( U_{\un{2}} \right)^{ba} \label{H1_ev} \\
    &\;\;\;\;- \frac{C_F}{N_c^2} \,  \tord \tr \left[ V_{\un{0}} \,  V_{\un{1}}^{\textrm{T} \dagger} \right] + \frac{1}{N_c^2} \,  \tord \tr \left[ V_{\un{0}}^{\dagger} t^b V_{\un{1}}^{\textrm{T}} t^a \right] \, \left( U_{\un{2}} \right)^{ba} - \frac{C_F}{N_c^2} \,  \tord \tr \left[ V_{\un{0}}^{\dagger} \,  V_{\un{1}}^{\textrm{T}} \right] \rrangle_2 (z')  \nonumber \\
    &\;\;\;\;+ \frac{\alpha_s N_c}{2\pi^2} \int\limits_{\frac{\Lambda^2}{s}}^z \frac{\dd{z}'}{z'} \int  \frac{\dd[2]{x}_2}{x_{21}^2} \, \textrm{Re} \, \llangle \frac{1}{N_c^2}   \tord \tr \Big[ t^b V_{\un{0}} t^a V_{\un{2}}^{\textrm{T} \dagger} \Big] \, U_{\un{1}}^{ba} + \frac{1}{N_c^2}  \tord \tr \Big[  t^a V_{\un{0}}^{\dagger} t^b V_{\un{2}}^{\textrm{T}} \Big] \, U_{\un{1}}^{ab} \rrangle_2  (z')\, . \nonumber
\end{align}
We can simplify this in the large-$N_c$ limit and then take the linearized, DLA contribution using the SU$(N_c)$ Fierz identities,
\begin{subequations}
    \begin{align}
        \tr \left[ t^a M_1 t^a M_2 \right] = \frac{1}{2} \tr \left[ M_1 \right] \tr \left[ M_2 \right] - \frac{1}{2N_c} \tr \left[ M_1 M_2 \right] , \\
        \tr \left[ t^a M_1 \right] \tr \left[ t^a M_2 \right] = \frac{1}{2} \tr \left[ M_1 M_2 \right] - \frac{1}{2N_c} \tr \left[ M_1 \right] \tr \left[ M_2 \right], 
    \end{align}    
\end{subequations}
for $N_c \times N_c$  matrices, $M_1$ and $M_2$, together with the relation between adjoint representation and fundamental representation eikonal Wilson lines,
\begin{align}
    U_{\un{2}}^{ba} = 2 \tr \left[ t^a V_{\un{2}}^{\dagger} t^b V_{\un{2}} \right] ,
\end{align}
as well as the fact that the correlators of products of dipole amplitudes factorizes in the large-$N_c$, mean-field limit into the product of the correlators,
\begin{align}
    \Big< \tr \left[ V_{\un{0}} V_{\un{1}}^{\textrm{pol} \ \dagger} \right] \tr \left[ V_{\un{0}} V_{\un{2}}^{\dagger} \right] \Big> \xrightarrow[\textrm{large}-N_c]{} \Big< \tr \left[ V_{\un{0}} V_{\un{1}}^{\textrm{pol} \ \dagger} \right] \Big> \Big< \tr \left[ V_{\un{0}} V_{\un{2}}^{\dagger} \right] \Big> .
\end{align}
Using these large-$N_c$ simplifications and linearizing the equation by setting eikonal dipole amplitudes to unity, $S_{10} \to 1$, yields
\begin{align}\label{trans_sing_ev}
    H_{10}^{1 T, S} (z) = H^{1 T, S \, (0)}_{10}(z) &+ \frac{\alpha_s N_c}{2\pi^2} \int\limits_{\frac{\Lambda^2}{s}}^z \frac{\dd{z}'}{z'} \int\limits_{\Lambda_{UV}}^{\Lambda_{IR}} \dd[2]{x}_2 \frac{x_{10}^2}{x_{21}^2 x_{20}^2}  \, \Big[ H_{12}^{1 T, S}(z') - \Gamma^{1 T, S}_{10, 21}(z') \Big] \\
    &+ \frac{\alpha_s N_c}{2\pi^2} \int\limits_{\frac{\Lambda^2}{s}}^z \frac{\dd{z}'}{z'} \int\limits_{\Lambda_{UV}}^{\Lambda_{IR}}  \frac{\dd[2]{x}_2}{x_{21}^2} \,  H_{21}^{1 T, S}(z') \, .   \notag
\end{align}
Here, as usual, $\Gamma^{1 T, S}$ is the neighbor dipole amplitude corresponding to $H^{1T,S}$ with physical transverse separation $x_{10}$ but whose lifetime is dictated by transverse size $x_{21}$. In Eq.~\eqref{trans_sing_ev}, we treat the UV and IR cutoffs similarly to the way they were handled in small-$x$ evolutions for other spin-dependent objects, that is, we take $\Lambda_{UV}$ to be the center-of-mass energy squared, $z's$, and we let $\Lambda_{IR}$ be dictated by the lifetime ordering condition, $x^2_{10}z \gg x^2_{21}z''$. Finally, as was done in previous cases, we impose an additional limit for the daughter dipole sizes from eikonal emissions to be smaller than those of the parents, such that $\frac{x^2_{10}}{x^2_{21}x^2_{20}}\to \frac{1}{x^2_{21}}\,\theta(x_{10}-x_{21})$. Altogether, the evolution equation~\eqref{trans_sing_ev}, together with its neighbor-dipole counterpart, becomes

\begin{subequations}\label{sub_lead_trans_ev}
\begin{tcolorbox}[ams align]
       H^{1 T, S} (x^2_{10},z) &= H^{1 T, S \, (0)}(x^2_{10},z) + \frac{\alpha_s N_c}{2\pi} \int\limits_{\frac{\Lambda^2}{s}}^z \frac{\dd{z}'}{z'} \int\limits_{1/z's}^{x_{10}^2 z/z'} \frac{\dd{x}_{21}^2}{x_{21}^2} \, H^{1 T, S}(x^2_{21},z') \label{sub_lead_trans_ev_bint_a} \\
      &\;\;\;\;+ \frac{\alpha_s N_c}{2\pi} \int\limits_{\max\{1/x^2_{10},\Lambda^2\}/s}^z \frac{\dd{z}'}{z'} \int\limits_{1/z's}^{x_{10}^2} \frac{\dd{x}_{21}^2}{x_{21}^2} \, \Big[ H^{1 T, S}(x^2_{21},z') - \Gamma^{1 T, S}(x^2_{10},x^2_{21},z') \Big] \, , \nonumber \\
      \Gamma^{1 T, S} (x^2_{10},x^2_{21},z') &= \Gamma^{1 T, S \, (0)}(x^2_{10},x^2_{21},z') + \frac{\alpha_s N_c}{2\pi} \int\limits_{\frac{\Lambda^2}{s}}^{z'} \frac{\dd{z}''}{z''} \int\limits_{1/z''s}^{x_{21}^2 z'/z''} \frac{\dd{x}_{32}^2}{x_{32}^2} \, H^{1 T, S}(x^2_{32},z'')  \label{sub_lead_trans_ev_bint_b} \\
     &\;\;\;\;+ \frac{\alpha_s N_c}{2\pi} \int\limits_{\max\{1/x^2_{10},\Lambda^2\}/s}^{z'} \frac{\dd{z}''}{z''} \int\limits_{1/z''s}^{\textrm{min}\{x_{10}^2,x_{21}^2 z'/z''\}} \frac{\dd{x}_{32}^2}{x_{32}^2} \, \Big[ H^{1 T, S}(x^2_{32},z'') - \Gamma^{1 T, S}(x^2_{10},x^2_{32},z'') \Big] \, . \nonumber
\end{tcolorbox}
\end{subequations}

\subsection{Asymptotic Solution}\label{sec:analytic_1TS}

In order to determine the high-energy asymptotic solution of Eqs.~\eqref{sub_lead_trans_ev}, 
we find it more convenient to work in the rescaled variables defined in \eqref{eta_s}. In terms of these new variables, the evolution equations become
\begin{subequations}\label{evol2}
\begin{align}
H^{1 T, S}(s_{10},\eta) = H^{1 T, S \, (0)}(s_{10},\eta) &+ \int\limits_{\max\{0,s_{10}\}}^{\eta}d\eta' \int\limits_{s_{10}}^{\eta'}ds_{21} \left[H^{1 T, S}(s_{21},\eta') - \Gamma^{1 T, S}(s_{10},s_{21},\eta')\right] \label{evol2_H} \\
&+ \int\limits_{0}^{\eta}d\eta' \int\limits_{s_{10}+\eta'-\eta}^{\eta'}ds_{21} \, H^{1 T, S}(s_{21},\eta')\,, \notag \\
\Gamma^{1 T, S}(s_{10},s_{21},\eta') = H^{1 T, S \, (0)}(s_{10},\eta') &+ \int\limits_{\max\{0,s_{10}\}}^{\eta'}d\eta''\int\limits_{\max\{s_{10},s_{21}+\eta''-\eta'\}}^{\eta''}ds_{32} \left[H^{1 T, S}(s_{32},\eta'') - \Gamma^{1 T, S}(s_{10},s_{32},\eta'')\right] \notag  \\
&+ \int\limits_{0}^{\eta'}d\eta'' \int\limits^{\eta''}_{s_{21}+\eta''-\eta'}ds_{32} \, H^{1 T, S}(s_{32},\eta'') \, . \label{evol2_Gamma}
\end{align}
\end{subequations}
From Eqs.~\eqref{evol2}, in order to know $H^{1 T, S}(s_{10},\eta)$ for some $0\leq s_{10}\leq \eta$, one needs to know $H^{1 T, S}(s_{21},\eta')$ in the parallelogram, $0\leq\eta'\leq\eta$ and $\eta'-\eta+s_{10}\leq s_{21}\leq \eta'$, which includes the negative-$s_{10}$ region. This is similar to the helicity problem at large $N_c\& N_f$ if we take $\Lambda$ to be an IR scale and not the IR cutoff \cite{Kovchegov:2020hgb,Adamiak:2023okq}. Furthermore, for $\Gamma^{1 T, S}(s_{10},s_{21},\eta')$, we always have $s_{21}\geq s_{10}$.

To proceed, we make another change of variable such that
\begin{align}\label{zeta_s}
&H^{1 T, S}(s_{10},\eta)\to H^{1 T, S}(\zeta = \eta-s_{10}, \eta)\;\;\;\;\;\text{and}\;\;\;\;\;\Gamma^{1 T, S}(s_{10},s_{21},\eta')\to\Gamma^{1 T, S}(s_{10},\zeta'=\eta'-s_{21},\eta')\,,
\end{align}
which turns Eqs.~\eqref{evol2} into
\begin{subequations}\label{evol3}
\begin{align}
H^{1 T, S}(\zeta,\eta) = H^{1 T, S \, (0)}(\zeta,\eta) &+ \int\limits_{0}^{\zeta}d\xi  \int\limits_{\max\{0,\eta-\zeta+\xi\}}^{\eta}d\eta' \left[H^{1 T, S}(\xi,\eta') - \Gamma^{1 T, S}(s_{10},\xi,\eta')\right] \label{evol3_H} \\
&+ \int\limits_{0}^{\zeta}d\xi \int\limits_{0}^{\eta}d\eta'  \, H^{1 T, S}(\xi,\eta')\,, \notag \\
\Gamma^{1 T, S}(s_{10},\zeta',\eta') = H^{1 T, S \, (0)}(\eta'-s_{10},\eta') &+ \int\limits_0^{\zeta'}d\xi \int\limits_{\max\{0,\xi+s_{10}\}}^{\eta'}d\eta'' \left[H^{1 T, S}(\xi,\eta'') - \Gamma^{1 T, S}(s_{10},\xi,\eta'')\right]  \label{evol3_Gamma} \\
&+ \int\limits_0^{\zeta'}d\xi \int\limits_0^{\eta'}d\eta'' \, H^{1 T, S}(\xi,\eta'') \,  .  \notag
\end{align}
\end{subequations}
First, we take the derivative of Eq.~\eqref{evol3_Gamma} with respect to $\eta'$ and obtain
\begin{align}\label{d_Gamma1}
    \frac{\partial}{\partial\eta'}\Gamma^{1 T, S}(s_{10},\zeta',\eta') &= \int\limits_0^{\zeta'}d\xi \left[2H^{1 T, S}(\xi,\eta') - \Gamma^{1 T, S}(s_{10},\xi,\eta')\right] .
\end{align}
Furthermore, we take another derivative of Eq.~\eqref{d_Gamma1} with respect to $\zeta'$ to get
\begin{align}\label{dd_Gamma}
    \frac{\partial^2}{\partial\zeta'\partial\eta'}\Gamma^{1 T, S}(s_{10},\zeta',\eta') &= 2H^{1 T, S}(\zeta',\eta') - \Gamma^{1 T, S}(s_{10},\zeta',\eta')\, .
\end{align}
If we put $s_{10}=\eta'-\zeta'$ in Eq.~\eqref{dd_Gamma}, then we obtain
\begin{align}\label{dd_Gamma_H}
    &\frac{\partial^2}{\partial\zeta'\partial\eta'}H^{1 T, S}(\zeta',\eta') = H^{1 T, S}(\zeta',\eta')\,,
\end{align}
which is solved by
\begin{align}\label{Laplace_H}
    &H^{1 T, S}(\zeta',\eta') = \int\frac{d\omega}{2\pi i}\,e^{\omega\eta' + \frac{\zeta'}{\omega}} h_{\omega}\,.
\end{align}
As the next step, we take yet another derivative of Eq.~\eqref{dd_Gamma} with respect to $s_{10}$ and get
\begin{align}\label{ddd_Gamma}
    \frac{\partial^3}{\partial s_{10}\partial\zeta'\partial\eta'}\Gamma^{1 T, S}(s_{10},\zeta',\eta') &= - \frac{\partial}{\partial s_{10}}\Gamma^{1 T, S}(s_{10},\zeta',\eta')\,.
\end{align}
With the help of Eq.~\eqref{Laplace_H}, the general solution of Eq.~\eqref{ddd_Gamma} takes the form of
\begin{align}\label{Laplace_Gamma}
    \Gamma^{1 T, S}(s_{10},\zeta',\eta') &= \int\frac{d\omega}{2\pi i}\,e^{\omega\eta' + \frac{\zeta'}{\omega}} h_{\omega} + \int\frac{d\omega}{2\pi i}\,e^{\omega\eta' - \frac{\zeta'}{\omega}} \left[g_{\omega}(s_{10}) - g_{\omega}(\eta'-\zeta')\right] .
\end{align}
Then, we plug Eqs.~\eqref{Laplace_H} and \eqref{Laplace_Gamma} into Eq.~\eqref{evol3_Gamma} to get
\begin{align}\label{evol4}
\int\frac{d\omega}{2\pi i}\,e^{\omega\eta' - \frac{\zeta'}{\omega}} \left[g_{\omega}(s_{10}) - g_{\omega}(\eta'-\zeta')\right] &= H^{1 T, S \, (0)}(\eta'-s_{10},\eta') + \int\frac{d\omega}{2\pi i}\left[1 - e^{\omega\eta'} - e^{\frac{\zeta'}{\omega}} \right]h_{\omega}   \\
&\;\;\;\;-  \int\limits_0^{\zeta'}d\xi \int\limits_{\max\{0,\xi+s_{10}\}}^{\eta'}d\eta'' \int\frac{d\omega}{2\pi i}\,e^{\omega\eta'' - \frac{\xi}{\omega}} \left[g_{\omega}(s_{10}) - g_{\omega}(\eta''-\xi)\right]  . \notag 
\end{align}
Before we proceed, we make use of Eq.~\eqref{Laplace_H} to see that
\begin{align}\label{inhom_evol4}
    &H^{1 T, S \, (0)}(\eta'-s_{10},\eta') + \int\frac{d\omega}{2\pi i}\left[1 - e^{\omega\eta'} - e^{\frac{\zeta'}{\omega}} \right]h_{\omega} \\
    &= H^{1 T, S \, (0)}(\eta'-s_{10},\eta') + H^{1 T, S \, (0)}(0,0) - H^{1 T, S \, (0)}(0,\eta') - H^{1 T, S \, (0)}(\zeta',0) \, , \notag
\end{align}
since $H^{1 T, S}$ is simply equal to its inhomogeneous term if either $\eta'=0$ or $\zeta'=0$. For brevity of the solution, we take the inhomogeneous term to be a constant from this point on: $H^{1 T, S \, (0)}(\zeta,\eta)\to H^{1 T, S \, (0)}$. As will be evident later, the final asymptotic solution will dominate any functional form of the inhomogeneous term. The most immediate consequence is that Eq.~\eqref{inhom_evol4} vanishes, leaving us with
\begin{align}\label{evol5}
\int\frac{d\omega}{2\pi i}\,e^{\omega\eta' - \frac{\zeta'}{\omega}} \left[g_{\omega}(s_{10}) - g_{\omega}(\eta'-\zeta')\right] &= -  \int\limits_0^{\zeta'}d\xi \int\limits_{\max\{0,\xi+s_{10}\}}^{\eta'}d\eta'' \int\frac{d\omega}{2\pi i}\,e^{\omega\eta'' - \frac{\xi}{\omega}} \left[g_{\omega}(s_{10}) - g_{\omega}(\eta''-\xi)\right]  . 
\end{align}
The integration limits in the right-hand side of Eq.~\eqref{evol5} require us to consider three separate cases: (i) $s_{10}\geq 0$, (ii) $-\zeta'\leq s_{10} < 0$ and (iii) $s_{10} < -\zeta'$.

\begin{enumerate}
    \item[(i)] $s_{10}\geq 0$ 

    In this case, since $\xi\geq 0$ in its range of integration, we always have $\xi+s_{10}\geq 0$. Hence, Eq.~\eqref{evol5} simplifies to
    \begin{align}\label{evol5_i}
        \int\frac{d\omega}{2\pi i}\,e^{\omega\eta' - \frac{\zeta'}{\omega}} \left[g_{\omega}(s_{10}) - g_{\omega}(\eta'-\zeta')\right] &= -  \int\limits_0^{\zeta'}d\xi \int\limits_{\xi+s_{10}}^{\eta'}d\eta'' \int\frac{d\omega}{2\pi i}\,e^{\omega\eta'' - \frac{\xi}{\omega}} \left[g_{\omega}(s_{10}) - g_{\omega}(\eta''-\xi)\right]  . 
    \end{align}
    Evaluating the integrals on the right-hand side yields
    \begin{align}\label{evol5_i1}
        &\int\frac{d\omega}{2\pi i} \left[ e^{\omega\eta'}g_{\omega}(s_{10}) - e^{\omega\eta'}g_{\omega}(\eta') + \int\limits_0^{\zeta'}d\xi \, e^{\omega\eta' - \frac{\xi}{\omega}}g'_{\omega}(\eta'-\xi) + \int\limits_0^{\zeta'}d\xi \int\limits_{\xi+s_{10}}^{\eta'}d\eta''\,e^{\omega\eta'' - \frac{\xi}{\omega}}\frac{1}{\omega} \, g'_{\omega}(\eta''-\xi) \right] =0 \,  ,
    \end{align}
    where we employed the following results from by-part integration,
    \begin{subequations}\label{by_part_1}
        \begin{align}
            \int\limits_{\xi+s_{10}}^{\eta'}d\eta'' \,e^{\omega\eta''} g_{\omega}(\eta''-\xi) &= \frac{1}{\omega} \, e^{\omega\eta'}g_{\omega}(\eta'-\xi) - \frac{1}{\omega} \, e^{\omega \left(\xi+s_{10}\right)}g_{\omega}(s_{10}) - \int\limits_{\xi+s_{10}}^{\eta'}d\eta''\,e^{\omega\eta''}\frac{1}{\omega} \, g'_{\omega}(\eta''-\xi) \, , \label{by_part_1a} \\
            \int\limits_0^{\zeta'}d\xi \, e^{- \frac{\xi}{\omega}}g_{\omega}(\eta'-\xi) &= \omega g_{\omega}(\eta') - e^{- \frac{\zeta'}{\omega}} \omega g_{\omega}(\eta'-\zeta') - \int\limits_0^{\zeta'}d\xi \, e^{- \frac{\xi}{\omega}} \omega g'_{\omega}(\eta'-\xi)  \, . \label{by_part_1b}
        \end{align}
    \end{subequations}
    As the next step, we take the derivative of Eq.~\eqref{evol5_i1} with respect to $s_{10}$ and get
    \begin{align}\label{evol5_i2}
        &\int\frac{d\omega}{2\pi i} \left[ e^{\omega\eta'}g'_{\omega}(s_{10}) - e^{\left(\omega - \frac{1}{\omega}\right)\zeta' + \omega s_{10}}\frac{1}{\omega^2-1} \, g'_{\omega}(s_{10}) + e^{\omega s_{10}}\frac{1}{\omega^2-1} \, g'_{\omega}(s_{10}) \right]  = 0 \, .
    \end{align}
    Then, we consider the $\zeta'=0$ case of Eq.~\eqref{evol5_i2}. Its $N$-th derivative with respect to $\eta'$, for any $N\geq 0$, is
    \begin{align}\label{evol5_i3}
        &\int\frac{d\omega}{2\pi i} \, e^{\omega\eta'}\omega^Ng'_{\omega}(s_{10}) = 0 \, .
    \end{align}
    To proceed, we write $g_{\omega}(s_{10})$ in term of its Laurent series,
    \begin{align}\label{g_Laurent}
        &g_{\omega}(s_{10}) = \sum_{n=-\infty}^{\infty}a_n(s_{10})\, \omega^n \, .
    \end{align}
    Then, Eq.~\eqref{evol5_i3} can be written as
    \begin{align}\label{evol5_i4}
        &\sum_{n=N+1}^{\infty}a'_{-n}(s_{10})\,\frac{\eta'^{n-N-1}}{(n-N-1)!} = 0 \, .
    \end{align}
    This forms countably many linear equations that inductively solve to $a'_{-n}(s_{10})=0$ for each $n\geq 1$. Altogether, this implies that
    \begin{align}\label{evol5_i5}
        &\int\frac{d\omega}{2\pi i}  \, e^{\omega\eta'}\left[g_{\omega}(s_{10}) - g_{\omega}(\eta')\right] = 0 
    \end{align}
    because all the contributing poles vanish as $a_{-n}(s_{10})=a_{-n}(\eta')$ for $n\geq 1$. Then, Eq.~\eqref{evol5_i1} simplifies to
    \begin{align}\label{evol5_i6}
        &\int\frac{d\omega}{2\pi i} \int\limits_0^{\zeta'}d\xi \left[ e^{\omega\eta' - \frac{\xi}{\omega}}g'_{\omega}(\eta'-\xi) + \int\limits_{\xi+s_{10}}^{\eta'}d\eta''\,e^{\omega\eta'' - \frac{\xi}{\omega}}\frac{1}{\omega} \, g'_{\omega}(\eta''-\xi) \right] =0 \,  .
    \end{align}
    Then, we take a derivative of Eq.~\eqref{evol5_i6} with respect to $s_{10}$, followed by $M$ derivatives with respect to $\zeta'$, for $M\geq 1$. At the end, we put $\zeta'=0$ in the result to obtain
    \begin{align}\label{evol5_i7}
        &0 = \int\frac{d\omega}{2\pi i} \, e^{\omega s_{10}}\frac{(\omega^2-1)^{M-1}}{\omega^M} \, g'_{\omega}(s_{10}) = \sum_{n=0}^{M-1}a'_n(s_{10}) \, \frac{1}{(M-n-1)!} \frac{\partial^{M-n-1}}{\partial\omega^{M-n-1}} \left[e^{\omega s_{10}} (\omega^2-1)^{M-1}\right]\Big|_{\omega=0} \,  .
    \end{align}
    The $M=1$ case gives $a'_0(s_{10})=0$, which together with the $M=2$ case yields $a'_1(s_{10})=0$. Inductively, one can show that $a'_n(s_{10})=0$ for each $n\geq 0$. Together with the previous similar results for $n\leq -1$, we conclude that $g'_{\omega}(s_{10})=0$, and so $g_{\omega}(s_{10})$ is independent of $s_{10}$. In particular, we have that $g_{\omega}(s_{10})-g_{\omega}(\eta'-\zeta')=0$.
        
    \item[(ii)] $-\zeta'\leq s_{10} < 0$

    In this case, the right-hand side of Eq.~\eqref{evol5} splits into two distinct regions, such that
    \begin{align}\label{evol5_ii}
        \int\frac{d\omega}{2\pi i}\,e^{\omega\eta' - \frac{\zeta'}{\omega}} \left[g_{\omega}(s_{10}) - g_{\omega}(\eta'-\zeta')\right] &= -  \int\limits_0^{-s_{10}}d\xi \int\limits_{0}^{\eta'}d\eta'' \int\frac{d\omega}{2\pi i}\,e^{\omega\eta'' - \frac{\xi}{\omega}} \left[g_{\omega}(s_{10}) - g_{\omega}(\eta''-\xi)\right]   \\
        &\;\;\;\;-  \int\limits_{-s_{10}}^{\zeta'}d\xi \int\limits_{\xi+s_{10}}^{\eta'}d\eta'' \int\frac{d\omega}{2\pi i}\,e^{\omega\eta'' - \frac{\xi}{\omega}} \left[g_{\omega}(s_{10}) - g_{\omega}(\eta''-\xi)\right]  . \notag 
    \end{align}
    Evaluating the integrals in the right-hand side via by-part integration similar to those in Eqs.~\eqref{by_part_1} yields
    \begin{align}\label{evol5_ii1}
        0 &= \int\frac{d\omega}{2\pi i}\left[g_{\omega}(s_{10}) - g_{\omega}(0) - e^{\omega\eta'} \left[g_{\omega}(s_{10}) - g_{\omega}(\eta')\right] + \int\limits_0^{-s_{10}}d\xi \, e^{-\frac{\xi}{\omega}} g'_{\omega}(-\xi) -  \int\limits_{0}^{\zeta'}d\xi \, e^{\omega\eta'-\frac{\xi}{\omega}} g'_{\omega}(\eta'-\xi) \right. \\
        &\;\;\;\;- \left. \int\limits_0^{-s_{10}}d\xi \int\limits_{0}^{\eta'}d\eta''\,e^{\omega\eta'' - \frac{\xi}{\omega}} \frac{1}{\omega} \, g'_{\omega}(\eta''-\xi) - \int\limits_{-s_{10}}^{\zeta'}d\xi \int\limits_{\xi+s_{10}}^{\eta'}d\eta''\,e^{\omega\eta'' - \frac{\xi}{\omega}} \frac{1}{\omega} \, g'_{\omega}(\eta''-\xi)  \right]  . \notag
    \end{align}
    Then, we take the derivative of Eq.~\eqref{evol5_ii1} with respect to $s_{10}$ and obtain
    \begin{align}\label{evol5_ii2}
        \int\frac{d\omega}{2\pi i}\left[1 - e^{\omega\eta'} \right] g'_{\omega}(s_{10})  + \int\frac{d\omega}{2\pi i} \,e^{\left(\omega-\frac{1}{\omega}\right)\zeta' + \omega s_{10}} \frac{1}{\omega^2-1} \, g'_{\omega}(s_{10}) - \int\frac{d\omega}{2\pi i} \,e^{\frac{s_{10}}{\omega}} \frac{\omega^2}{\omega^2-1} \, g'_{\omega}(s_{10}) &= 0 \,  .  
    \end{align}
    First, its $N$-th derivative with respect to $\eta'$, for any $N\geq 1$, gives 
    \begin{align}\label{evol5_ii3}
        &\int\frac{d\omega}{2\pi i} \, e^{\omega\eta'}\omega^Ng'_{\omega}(s_{10}) = 0 \, ,
    \end{align}
    which is exactly Eq.~\eqref{evol5_i3} but in this case its applicability does not include the $N=0$ case. In turn, this implies that $a'_{-n}(s_{10})=0$ for each $n\geq 2$, as we write $g_{\omega}(s_{10})$ in term of its Laurent series, c.f. Eq.~\eqref{g_Laurent}. 
    
    To proceed, we revisit Eq.~\eqref{evol5_ii2} and take its $M$-th derivative with respect to $\zeta'$, for some $M\geq 1$. Then, its $\zeta'=0$ case gives
    \begin{align}\label{evol5_ii4}
        0 &= \int\frac{d\omega}{2\pi i} \, e^{\omega s_{10}} \frac{(\omega^2-1)^{M-1}}{\omega^M}\,g'_{\omega}(s_{10}) = \sum_{n=-1}^{M-1}a'_n(s_{10}) \, \frac{1}{(M-n-1)!} \frac{\partial^{M-n-1}}{\partial\omega^{M-n-1}} \left[e^{\omega s_{10}} (\omega^2-1)^{M-1}\right]\Big|_{\omega=0} \,  . 
    \end{align}
    Note that this result is different from the previous case because $a'_{-1}(s_{10})$ can still be nonzero. As a result, we instead obtain a recipe to inductively write $a'_n(s_{10})$ in term of $a'_{-1}(s_{10})$ multiplied by a factor that depends on $s_{10}$. Specifically, we have
    \begin{align}\label{evol5_ii5}
        a'_n(s_{10}) = \begin{cases}
            (-1)^{n+1}\left[\frac{s_{10}^{n-1}}{(n-1)!} + \frac{s_{10}^{n+1}}{(n+1)!}\right] a'_{-1}(s_{10})&,\;\;\; n\geq 1 \\
            -s_{10}\,a'_{-1}(s_{10})&,\;\;\; n=0
        \end{cases} \; .
    \end{align}
    This leads to the general form,
    \begin{align}\label{evol5_ii6}
        g'_{\omega}(s_{10}) &= a'_{-1}(s_{10}) \left\{\frac{1}{\omega} - s_{10} + \sum_{n=1}^{\infty}(-1)^{n+1}\left[\frac{s_{10}^{n-1}}{(n-1)!} + \frac{s_{10}^{n+1}}{(n+1)!}\right]\omega^n\right\}  = a'_{-1}(s_{10}) \left(\omega+\frac{1}{\omega}\right) e^{-\omega s_{10}} \, .  
    \end{align}
    As the next step, we will plug this result into Eq.~\eqref{evol5_ii1}. To do so, we first notice that
    \begin{align}\label{evol5_ii7}
        \int\frac{d\omega}{2\pi i} \, e^{\omega t}\left[g_{\omega}(s_{10}) - g_{\omega}(t) \right] &= \int\frac{d\omega}{2\pi i} \, e^{\omega t} \int\limits_t^{s_{10}}ds \, g'_{\omega}(s) = \int\limits_t^{s_{10}}ds \, a'_{-1}(s) \int\frac{d\omega}{2\pi i} \, e^{\omega (t-s)}\left(\omega+\frac{1}{\omega}\right) \\
        &= \int\limits_t^{s_{10}}ds \, a'_{-1}(s) = a_{-1}(s_{10}) - a_{-1}(t) \, , \notag
    \end{align}
    for $t\in\{0,\eta'\}$. Then, Eq.~\eqref{evol5_ii1} gives
    \begin{align}\label{evol5_ii8}
        &a_{-1}(\eta') - a_{-1}(0) + \int\limits_0^{-s_{10}}d\xi\,a'_{-1}(-\xi)\left[J_0(2\xi)+J_2(2\xi)\right] - \int\limits_0^{\zeta'}d\xi\,a'_{-1}(\eta'-\xi)\left[J_0(2\xi)+J_2(2\xi)\right] = 0 \, .
    \end{align}
    The derivative of Eq.~\eqref{evol5_ii8} with respect to $\zeta'$ yields
    \begin{align}\label{evol5_ii9}
        &a'_{-1}(\eta'-\zeta')\left[J_0(2\zeta')+J_2(2\zeta')\right] = 0 \, ,
    \end{align}
    for any $\eta'$ and $\zeta'$. The only way for this to hold is to have $a'_{-1}(s_{10})=0$. Hence, together with Eq.~\eqref{evol5_ii5} and the previous result for $a'_n$ with $n\leq -2$, we conclude that $g'_{\omega}(s_{10})=0$.

    \item[(iii)] $s_{10} < -\zeta'$

    In this case, we always have $\xi+s_{10}\leq \zeta'+s_{10} < 0$ on the right-hand side of Eq.~\eqref{evol5}. This allows us to express the latter as
    \begin{align}\label{evol5_iii}
        \int\frac{d\omega}{2\pi i}\,e^{\omega\eta' - \frac{\zeta'}{\omega}} \left[g_{\omega}(s_{10}) - g_{\omega}(\eta'-\zeta')\right] &= -  \int\limits_0^{\zeta'}d\xi \int\limits_{0}^{\eta'}d\eta'' \int\frac{d\omega}{2\pi i}\,e^{\omega\eta'' - \frac{\xi}{\omega}} \left[g_{\omega}(s_{10}) - g_{\omega}(\eta''-\xi)\right]  . 
    \end{align}
    Then, the derivative of Eq.~\eqref{evol5_iii} with respect to $s_{10}$ gives
    \begin{align}\label{evol5_iii1}
        \int\frac{d\omega}{2\pi i} \left[e^{\omega\eta'} + e^{- \frac{\zeta'}{\omega}} - 1 \right] g'_{\omega}(s_{10}) &= 0 \, , 
    \end{align}
    where we also evaluated the relatively straightforward integrals that remained. Then, the $N$-th derivative of Eq.~\eqref{evol5_iii1} with respect to $\eta'$ for any $N\geq 1$ yields exactly the same scenario as that of Eq.~\eqref{evol5_ii3}. Hence, we also conclude that $a'_{-n}(s_{10})=0$ for any $n\geq 2$ in the Laurent series of $g_{\omega}(s_{10})$ in the current case.

    Furthermore, we take the $M$-th derivative of Eq.~\eqref{evol5_iii1} with respect to $\zeta'$, for $M\geq 1$, and subsequently put $\zeta'=0$ to obtain
    \begin{align}\label{evol5_iii2}
        0 &= \int\frac{d\omega}{2\pi i} \, \frac{1}{\omega^M} g'_{\omega}(s_{10}) = \sum_{n=-1}^{N-1}a'_n(s_{10})  \,\frac{1}{(N-n-1)!}\, .
    \end{align}
    This forms a system of countably many linear equations that solves to
    \begin{align}\label{evol5_iii3}
        a'_n(s_{10}) &= \frac{(-1)^{n+1}}{(n+1)!}\,a'_{-1}(s_{10})\,,
    \end{align}
    for any $n\geq -1$. This implies 
    \begin{align}\label{evol5_iii4}
        g'_{\omega}(s_{10}) &= a'_{-1}(s_{10})\,\frac{1}{\omega}\,e^{-\omega}\,.
    \end{align}
    Finally, plugging this result back into Eq.~\eqref{evol5_iii1}, we obtain
    \begin{align}\label{evol5_iii5}
        a'_{-1}(s_{10})\,I_0\left(2\sqrt{\zeta'\,}\right) = 0 \, ,
    \end{align}
    which allows us to deduce that $a'_{-1}(s_{10})=0$ and hence $g'_{\omega}(s_{10})=0$.
    
\end{enumerate}

Because $g_{\omega}(s_{10})$ is independent of $s_{10}$ in all cases, we conclude in light of Eq.~\eqref{Laplace_Gamma} that 
\begin{align}\label{Gamma_H}
    \Gamma^{1 T, S}(s_{10},\zeta',\eta') &= H^{1 T, S}(\zeta',\eta') = \int\frac{d\omega}{2\pi i}\,e^{\omega\eta' + \frac{\zeta'}{\omega}} h_{\omega}\,. 
\end{align}
Physically, this means that the neighbor dipole behaves like the ordinary dipole whose transverse size is determined by the (harder) transverse scale that characterizes its lifetime. This allows us to rewrite Eq.~\eqref{evol3_H} as
\begin{align}\label{evol6}
    H^{1 T, S}(\zeta,\eta) = H^{1 T, S \, (0)}(\zeta,\eta) + \int\limits_{0}^{\zeta}d\xi \int\limits_{0}^{\eta}d\eta'  \, H^{1 T, S}(\xi,\eta')\,. 
\end{align}
This evolution equation is identical to the one for flavor-nonsinglet helicity distribution, which has been solved in \cite{Kovchegov:2016zex} to yield asymptotic solution,
\begin{align}\label{asymp_H}
    &H^{1 T, S}(\zeta,\eta) \sim e^{2\sqrt{\eta\zeta}} \, ,
\end{align}
at large $\eta\zeta = \frac{\alpha_sN_c}{2\pi} \, \ln\frac{zs}{\Lambda^2} \, \ln(zsx^2_{10})$. As promised at the beginning of this section, this asymptotic solution dominates the inhomogeneous term, which usually grows at most with logarithms of $zs$ and/or $x^2_{10}$. In terms of the TMDs, we have the small-$x$ asymptotic form,
\begin{align}\label{asymp_TMD_2}
    &h_{1T}^S(x,k^2_{\perp}) \sim h_{1T}^{\perp\,S}(x,k^2_{\perp}) \sim x\left(\frac{1}{x}\right)^{\sqrt{2\alpha_sN_c/\pi}} \, ,
\end{align}
c.f. Eqs.~\eqref{trans_singlet_simp} and \eqref{pretz_singlet_simp}. This implies that the flavor singlet transversity and pretzelosity TMDs have the same small-$x$ asymptotic behaviors as their flavor non-singlet counterparts \cite{Kovchegov:2018zeq,Santiago:2023rfl}. Note that this also agrees with the results for the flavor singlet transversity PDF as studied in the Infrared Evolution Equation Framework \cite{Kirschner:1996jj} where the evolution kernels were explicitly matched to the small-$x$ DLA limit of the transverse spin dependent Dokshitzer-Gribov-Lipatov-Altarelli-Parisi (DGLAP) evolution equations \cite{Artru:1989zv}, showing agreement between all three frameworks for the DLA small-$x$ asymptotics of the transversity PDF.

\subsection{Numerical Crosscheck}

In order to verify the analytic solution derived above, we perform an iterative computation of the dipole amplitude, $H^{1 T, S}$, together with its neighbor dipole counterpart, $\Gamma^{1 T, S}$, following the evolution equations~\eqref{evol2}. To do so, we follow the process akin to \cite{Kovchegov:2016weo,Kovchegov:2020hgb,Cougoulic:2022gbk,Adamiak:2023okq}, discretizing the $\eta$, $s_{10}$ and $s_{21}$ space with step size $\Delta$. Explicitly, the dipole amplitudes can be written as
\begin{align}\label{H_Gamma_disc}
    &H^{1 T, S}_{ij} = H^{1 T, S}(s_{10}=i\Delta, \eta=j\Delta)\;\;\;\;\;\text{and}\;\;\;\;\;\Gamma^{1 T, S}_{ikj} = \Gamma^{1 T, S}(s_{10}=i\Delta, s_{21}=k\Delta, \eta'=j\Delta) \, .
\end{align}
Typically, $i$, $j$ and $k$ take on integer values. Then, we discretize the integrals in the evolution equations~\eqref{evol2} based on the left-hand convention of Riemann sum. This gives
\begin{subequations}\label{disc2d}
    \begin{align}
        H^{1T,S}_{ij} &= H^{1T,S(0)}_{ij} + \Delta^2\sum_{j'=\max\{0,i\}}^{j-1}\sum_{i'=i}^{j'-1}\left[H_{i'j'} - \Gamma_{ii'j'}\right] + \Delta^2\sum_{j'=0}^{j-1}\sum_{i'=i+j'-j}^{j'-1}H_{i'j'} \, , \label{disc2d_H} \\
        \Gamma^{1T,S}_{ikj} &= H^{1T,S(0)}_{ij} + \Delta^2\sum_{j'=\max\{0,i\}}^{j-1}\sum_{i'=\max\{i,k+j'-j\}}^{j'-1}\left[H_{i'j'} - \Gamma_{ii'j'}\right] + \Delta^2\sum_{j'=0}^{j-1}\sum_{i'=k+j'-j}^{j'-1}H_{i'j'} \, . \label{disc2d_Gamma}
    \end{align}
\end{subequations}
From the summations, we see that $H^{1T,S}_{ij}=H^{1T,S(0)}_{ij}$ whenever $j=0$ or $i=j$. Similarly, $\Gamma^{1T,S}_{ikj}=H^{1T,S(0)}_{ij}$ if $j=0$ or $i=j$, as the latter condition would force $i=k=j$. Finally, we recall that $\Gamma^{1T,S}_{iij}=H^{1T,S}_{ij}$. Besides these edge cases, we have the recursive relations,
\begin{subequations}\label{disc_recursive}
    \begin{align}
        H^{1T,S}_{ij} &= H^{1T,S(0)}_{ij} - H^{1T,S(0)}_{i(j-1)} + H^{1T,S}_{i(j-1)} + \Delta^2\sum_{j'=0}^{j-2} H_{(i+j'-j)j'} + \Delta^2\sum_{i'=i-1}^{j-2} H_{i'(j-1)} + \Delta^2\sum_{i'=i}^{j-2}\left[H_{i'(j-1)} - \Gamma_{ii'(j-1)}\right] , \label{disc_recursive_H} \\
        \Gamma^{1T,S}_{ikj} &= H^{1T,S(0)}_{ij} - H^{1T,S(0)}_{i(j-1)} + \Gamma^{1T,S}_{i(k-1)(j-1)} + \Delta^2\sum_{i'=k-1}^{j-2}\left[2H_{i'(j-1)} - \Gamma_{ii'(j-1)}\right] . \label{disc_recursive_Gamma}
    \end{align}
\end{subequations}

\begin{figure*}[t!]
    \centering
    \begin{subfigure}[t!]{0.44\textwidth}
        \centering
        \includegraphics[width=\textwidth]{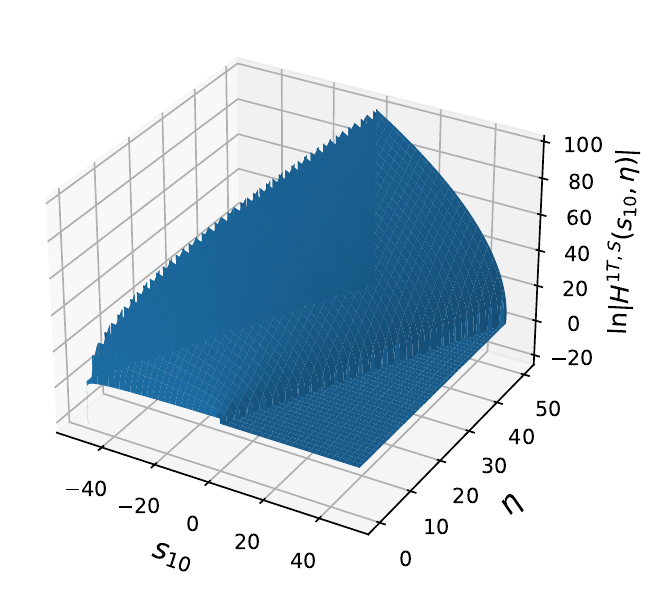}
        \caption{Three-dimensional plot of $\ln\left|H^{1T,S}(s_{10},\eta)\right|$ versus $\eta$ and $s_{10}$}
        \label{fig:H1TS_1D}
    \end{subfigure}%
    ~ 
    \begin{subfigure}[t!]{0.4\textwidth}
        \centering
        \includegraphics[width=\textwidth]{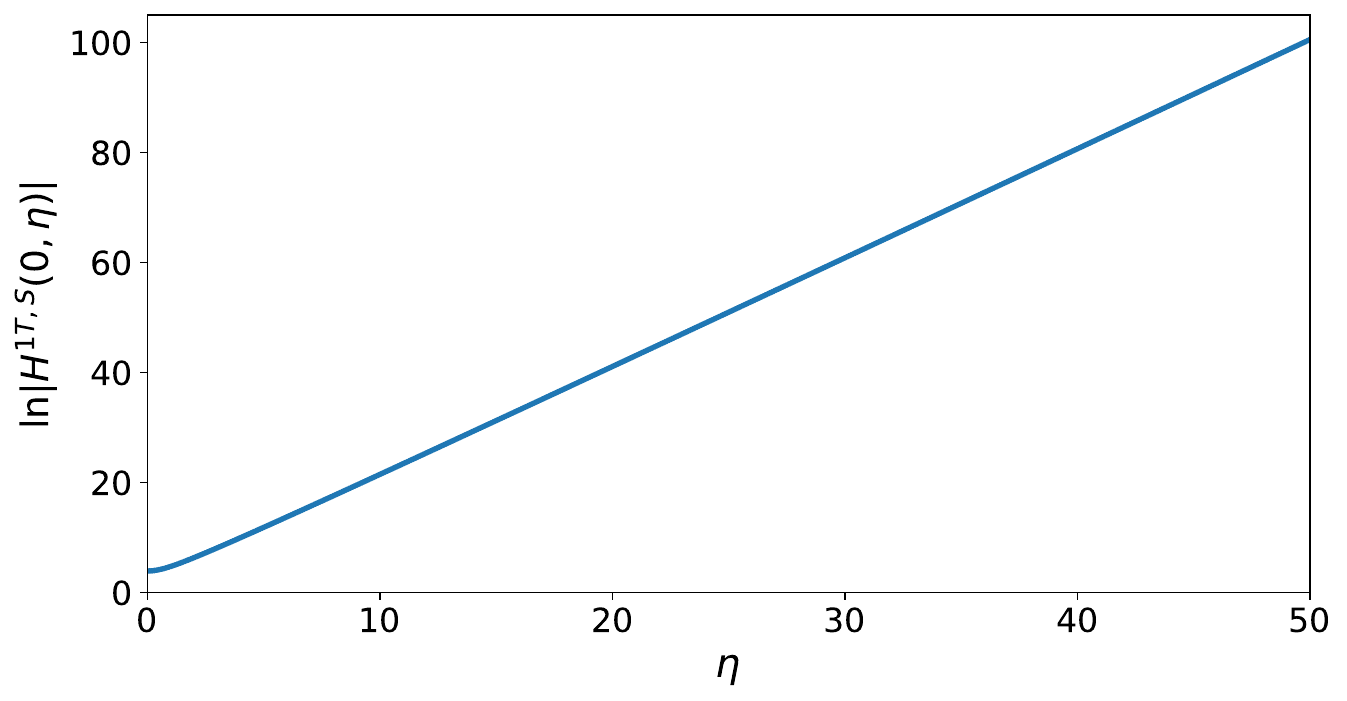}
        \caption{Two-dimensional plot of $\ln\left|H^{1T,S}(0,\eta)\right|$ against $\eta$}
        \label{fig:H1TS_2D}
    \end{subfigure}
    \caption{Plots illustrating $H^{1T,S}(s_{10},\eta)$ that results from initial condition~\eqref{ICs10}.}
    \label{fig:H1TS}
\end{figure*}

Based on Eqs.~\eqref{disc_recursive}, we iteratively compute the dipole amplitudes at each discrete point within $0\leq j\leq j_{\max}$ and $j-j_{\max}\leq i\leq k\leq j$, using $j_{\max}=500$ and step size $\Delta=0.1$. We start from the initial condition,
\begin{align}\label{ICs10}
    &H^{1T,S(0)}_{ij} = i\Delta + 50 =  s_{10} + 50\, ,
\end{align}
whose dependence on $s_{10}$ is important because otherwise $\Gamma^{1T,S}_{ikj}=H^{1T,S}_{ij}$ would have been trivially satisfied, exactly reducing Eqs.~\eqref{disc2d} to the discrete counterpart of Eq.~\eqref{evol6}. The results are plotted in Figs.~\ref{fig:H1TS}, including a three-dimensional plot of $\ln\left|H^{1T,S}(s_{10},\eta)\right|$ and a two-dimensional plot of $\ln\left|H^{1T,S}(0,\eta)\right|$. The latter highlights the exponential asymptotic behavior, $H^{1T,S}(0,\eta)\sim e^{\alpha_{1T,S}\eta}$. A linear regression on the top 25\% of the data points in Fig.~\ref{fig:H1TS_2D} yields the intercept of $\hat{\alpha}_{1T,S} = 1.986$, which is close to the analytic solution, $\alpha_{1T,S}=2$, which can be obtained directly from Eq.~\eqref{asymp_H} along $\zeta=\eta$ line. Although one would need multiple such runs with different step and grid sizes to quantify the uncertainty of the intercept prediction, the result from this grid serves as a confirmation of the analytic solution derived in Sec.~\ref{sec:analytic_1TS}, and the resulting small-$x$ asymptotics for the flavor singlet transversity and pretzelosity TMDs.

Additionally, it is informative to compare the small-$x$ evolution equations~\eqref{sub_lead_trans_ev} for these flavor singlet TMDs to that of the non-singlet counterparts~\cite{Santiago:2023rfl}, the latter of which reads
\begin{align}\label{trans_pretz_NS_evol}
    H^{1T,NS}(x^2_{10},z) &= H^{1T,NS\,(0)}(x^2_{10},z) + \frac{\as N_c}{2\pi}\int\limits_{\Lambda^2/s}^z\frac{dz'}{z'}\int\limits_{1/z's}^{x^2_{10}z/z'}\frac{dx^2_{21}}{x^2_{21}} \, H^{1T,NS}(x^2_{21},z')\,.
\end{align}
Mathematically, Eq.~\eqref{trans_pretz_NS_evol} is identical to the first line of Eq.~\eqref{sub_lead_trans_ev_bint_a}, that is, without the term proportional to $H^{1T,S}-\Gamma^{1T,S}$. Although the singlet and non-singlet evolution equations are completely different, as they were constructed from separate calculations via totally different amplitudes~\footnote{To see this explicitly, compare Eqs.~\eqref{trans_singlet_simp}-\eqref{eqn:54} for the singlet case to Eqs.~(35), (44)-(45) and (47) in~\cite{Santiago:2023rfl} for the non-singlet case.}, they result in the same large-rapidity asymptotic behaviors for the respective amplitudes because the term proportional to $H^{1T,S}-\Gamma^{1T,S}$ in the singlet evolution equation yields a term asymptotically suppressed per the discussion in Section~\ref{sec:analytic_1TS}. In particular, extra care needs to be taken in order to completely understand the subtle but complete difference between the two evolution equations despite the similarity of their asymptotic solutions.

\begin{figure*}[t!]
    \centering
    \begin{subfigure}[t!]{0.42\textwidth}
        \centering
        \includegraphics[width=\textwidth]{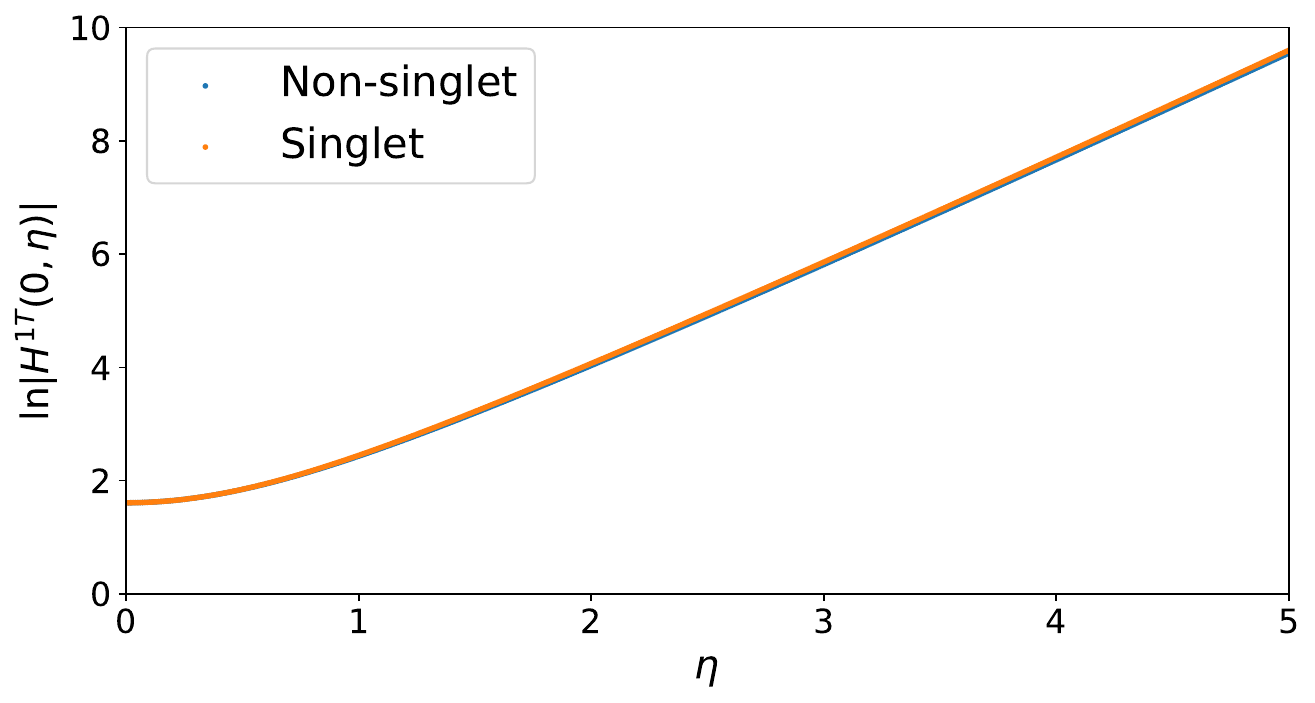}
        \caption{Plots of $\ln|H^{1T,NS}(0,\eta)|$ (blue) and $\ln|H^{1T,S}(0,\eta)|$ (orange) computed using $\Delta=0.01$}
        \label{fig:lnH_comp}
    \end{subfigure}%
    ~ 
    \begin{subfigure}[t!]{0.42\textwidth}
        \centering
        \includegraphics[width=\textwidth]{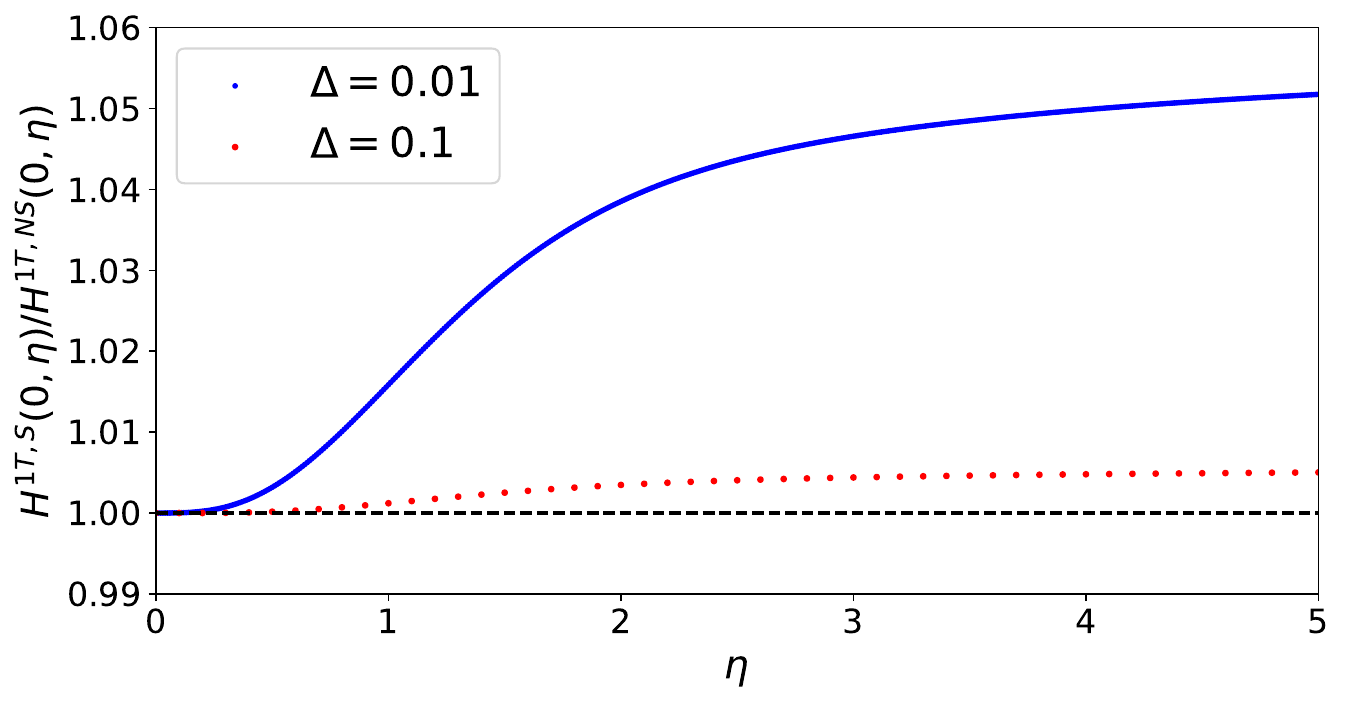}
        \caption{Plots of the ratios, $H^{1T,S}(0,\eta)/H^{1T,NS}(0,\eta)$, computed using $\Delta=0.01$ (blue) and $\Delta=0.1$ (red)}
        \label{fig:H_ratio}
    \end{subfigure}
    \caption{Plots illustrating the differences in solutions of the evolution equations for the flavor singlet and non-singlet transversity and pretzelosity TMDs.}
    \label{fig:H1TS}
\end{figure*}

To further illustrate the similarity in their solutions' asymptotic behaviors, we compute the numerical solutions to both equations with step size, $\Delta=0.01$. Fig.~\ref{fig:lnH_comp} shows the plots of $\ln|H^{1T,NS}(0,\eta)|$ and $\ln|H^{1T,S}(0,\eta)|$ -- corresponding respectively to the non-singlet and singlet equations -- that result from the numerical computations. We see that the difference in truly minimal in the logarithmic scale where only the most dominant asymptotic behavior is highlighted. To see more clearly the difference between the two solutions, we show in Fig.~\ref{fig:H_ratio} the ratio, $H^{1T,S}(0,\eta)/H^{1T,NS}(0,\eta)$, computed using step sizes, $\Delta=0.01$ and $\Delta=0.1$. Here, the ratio grows clearly distinct from 1, which would be the expectation had the two evolution equations are completely identical. A close observation of the behavior of this ratio along the $\eta$-axis implies that it increases from 1 at $\eta=0$ to an inflection point then keeps going up at ever decreasing rates as $\eta\to\infty$~\footnote{We see this trend up to $\eta=50$.}. However, the ratio at each given $\eta$ appears dependent on the step size, $\Delta$, employed in the numerical computation. As a result, the discretization steps involved in computing the numerical solutions should also play a significant role in Fig.~\ref{fig:H_ratio}. Nevertheless, the main takeaway of this exercise still stands, that the singlet and non-singlet evolution equations contain explicit, albeit subtle, differences in their solutions.

At $\eta=5$ in Fig.~\ref{fig:H_ratio}, we see 0.5\% difference for $\Delta=0.1$ and 5\% for $\Delta=0.01$. In the continuum limit, we have a larger difference between the singlet and non-singlet. It is expected that the statistical uncertainties for double-transverse spin-asymmetries measured at the future EIC will be smaller than $5\%$ \cite{Seidl:2022dhh,Seidl:2022dmh} Relative to EIC precision, it is important to use the correct evolution equation, particularly including the $H^{1T,S}-\Gamma^{1T,S}$ term into the singlet equation.


\section{ Boer-Mulders and worm-gear H TMDs} \label{sec:bmh}

Finally, we turn to the Boer-Mulders, $h_1^{\perp}$ and worm-gear $h_{1L}^{\perp}$ TMDs. The operator definition 
\begin{align}\label{bm_def}
- \frac{\epsilon^{ij} k^i}{M_P} \, h_1^{\perp \, q} (x, k_T^2) & \subset \int \frac{d r^- \, d^2 r_\perp}{2 (2\pi)^3} \, e^{i k \cdot r} \bra{P} \bar{\psi}(0) \mathcal{U}[0,r] \frac{ i \, \sigma^{j+} \, \gamma^5}{2} \psi(r) \ket{P} .
\end{align}
The expression in terms of polarized dipole amplitudes for the flavor singlet case is 

\begin{align}
\label{bm_singlet_simp}
     \frac{k^y}{M_P} \, h_1^{\perp \, \textrm{S}} (x, k_T^2) &= \frac{ixN_c}{2\pi^4} \int \frac{dz}{z} \int d^2x_0\, d^2x_1  \int\frac{d^2k_1}{(2\pi)^2}  \, e^{i(\kk+\kk_1)\cdot\xx_{10}} \,\frac{1}{k_{1\perp}^2k_{\perp}^2}\left(\frac{1}{k_{1\perp}^2} + \frac{1}{k_{\perp}^2}\right)       \notag \\
    &\;\;\;\;\;\times  \Big\{ - \left[2(\underline{S}\cdot\kk_1)(\underline{S}\cdot\kk) - (\kk_1\cdot\kk)\right] ({\un x}_{10} \times {\un S}) H^1(x_{10}^2, z) \\
    &+ \left[(\underline{S}\cdot\kk_1)(\underline{S}\times\kk) + (\underline{S}\times\kk_1)(\underline{S}\cdot\kk)\right] ({\un x}_{10} \vdot {\un S}) H^2(x_{10}^2, z) \Big\} \notag ,
\end{align}
where the polarized dipole amplitudes are
\begin{subequations}\label{bm_dips}
\begin{align}
& H^{1}_{10} (z) \equiv \frac{1}{2 N_c} \, \mbox{Im} \,  \llangle \tord \tr \left[ V_{\underline{0}} \, V^{\textrm{T} \, \dagger}_{{\un 1}} \right] - \tord \tr \left[ V_{\underline{0}}^\dagger \, V^{\textrm{T} }_{{\un 1}} \right] \rrangle_2, \\
& H^{2}_{10} (z) \equiv \frac{1}{2 N_c} \, \mbox{Re} \,  \llangle \tord \tr \left[ V_{\underline{0}} \, V^{\textrm{T} \, \perp \, \dagger}_{{\un 1}} \right] + \tord \tr \left[ V_{\underline{0}}^\dagger \, V^{\textrm{T} \, \perp}_{{\un 1}} \right]  \rrangle_2 , 
\end{align}
\end{subequations}
with polarized Wilson lines as defined in \eq{t_wlines}, which when integrated over impact parameter have the form
\begin{subequations}\label{bm_bint_dips}
\begin{align}
& \int d^2 b_\perp \, H^{1}_{10} (z) = ({\un x}_{10} \times {\un S}) \, H^{1} (x_{10}^2, z ) ,   \\
& \int d^2 b_\perp \, H^{2}_{10} (z) = ({\un x}_{10} \cdot {\un S}) \, H^{2} (x_{10}^2, z ) .
\end{align}
\end{subequations}

The operator definition for the worm-gear $h_{1L}^{\perp}$ TMD is
\begin{align} \label{sseh1l_1}
    \frac{k_T^x}{M_P} h_{1L}^{\perp} (x, k_T^2) &= \frac{1}{2} \sum_{S_P} S_P \int \frac{\dd{r}^- \dd[2]{r}_{\perp}}{2(2\pi)^3} \bra{P, S_P} \bpsi (0) \mathcal{U} [0, r] \frac{i \sigma^{1+}\gamma_5}{2} \psi(r)  \ket{P, S_P} \\
    &=\frac{1}{2} \sum_{S_P} S_P \int \frac{\dd{r}^- \dd[2]{r}_{\perp}}{2(2\pi)^3} \bra{P, S_P} \bpsi (0) \mathcal{U} [0, r] \frac{\gamma_5 \gamma^+ \gamma^1}{2} \psi(r)  \ket{P, S_P} \notag
\end{align}
where the proton spin $S_P$ is now in the longitudinal direction. The expression in terms of polarized dipole amplitudes for the flavor singlet TMD is 

\begin{align}
\label{sseh1l_3}
    &\frac{k_T^x}{M_P} h_{1L}^{\perp \ \textrm{S}} (x, k_T^2) = -  \frac{xN_c}{2\pi^4} \int \frac{dz}{z} \int d^2x_0 \,  d^2x_1  \int\frac{d^2k_1}{(2\pi)^2}  \,  e^{i(\kk+\kk_1)\cdot \xx_{10}} \, \frac{1}{k_{1\perp}^2k_{\perp}^2}\left(\frac{1}{k_{1\perp}^2} + \frac{1}{k_{\perp}^2}\right)     \\
    &\times \left\{ \left[2(\underline{S}\cdot\kk_1)(\underline{S}\cdot\kk) - (\kk_1\cdot\kk)\right] ({\un x}_{10} \times {\un S}) Hl^{1} (x_{10}^2, z )  +  \left[(\underline{S}\cdot\kk_1)(\underline{S}\times\kk) + (\underline{S}\times\kk_1)(\underline{S}\cdot\kk)\right] ({\un x}_{10} \vdot {\un S}) Hl^{2} (x_{10}^2, z ) \right\} , \notag
\end{align}
where the polarized dipole amplitudes are defined analogously to \eq{bm_bint_dips} in terms of the sub-sub-eikonal dipole amplitudes are given by
\begin{subequations}\label{HL}
\begin{align}
Hl^{1}_{10} (z ) &= \frac{1}{2N_c}\sum_f \, \mbox{Re}\, \llangle \tord \tr \left[ V_{\underline{0}} \, V^{\textrm{T} \, \dagger}_{{\un 1}} \right] - \tord \tr \left[ V_{\underline{0}}^\dagger \, V^{\textrm{T} }_{{\un 1}} \right] \rrangle_2 ,    \\
Hl^{2}_{10} (z ) &= \frac{1}{2N_c}\sum_f\,\mbox{Im}   \,  \llangle \tord \tr \left[ V_{\underline{0}} \, V^{\textrm{T} \, \perp \, \dagger}_{{\un 1}} \right] + \tord \tr \left[ V_{\underline{0}}^\dagger \, V^{\textrm{T} \, \perp}_{{\un 1}} \right]  \rrangle_2  .
\end{align}
\end{subequations}

The polarized dipole amplitudes entering \eq{bm_singlet_simp} and \eq{sseh1l_3} are essentially the same, so we can obtain the small-$x$ asymptotics of both the TMDs $h_1^{\perp}$ and $h_{1L}$ by solving the small-$x$ evolution equations for the asymptotics of a single set of polarized dipole amplitudes. Turning to the evolution of these dipole amplitudes, the linearity in $\un{x}_{10}$ of the impact parameter integrated dipole amplitudes in \eq{bm_bint_dips} tells us that the polarized quark emission diagrams for evolution will vanish. The polarized gluon emissions are still mass suppressed, so we only have eikonal emissions to consider. This is exactly what happened in the flavor non-singlet case, where the only change was a minus sign in the definitions in \eq{bm_dips} as opposed to the plus signs here. This sign change doesn't affect the eikonal exchanges, so we have the same evolution and asymptotics. Namely, the $\sqrt{\alpha_s N_c / 2 \pi}$ type corrections to the scaling with $x$ are washed out by the $z$ integral in \eq{bm_singlet_simp}, so we get 
\begin{align}
    h_1^{\perp \, S} (x \ll 1, k_T^2) \sim h_{1L}^{\perp \, S} (x \ll 1, k_T^2) \sim \left( \frac{1}{x} \right)^0 ,
\end{align}
finding that the flavor singlet Boer-Mulders TMD also has almost exactly sub-eikonal scaling even after linearized DLA evolution, while the worm-gear $h_{1L}^{\perp}$ has almost exactly sub-sub-eikonal scaling in the same approximation.


\section{Conclusions}\label{sec:conc}

We have applied the LCOT to study the small-$x$ asymptotics of the six leading-twist, flavor singlet quark TMDs, namely the Sivers function, $f_{1T}^{\perp}$, the transverse worm-gear, $g_{1T}$, the Boer-Mulders function, $h_1^{\perp}$, the helicity worm-gear, $h_{1L}$, the transversity, $h_{1T}$, and the pretzelosity, $h_{1T}^{\perp}$.

We have derived expressions for all six transverse momentum-dependent distributions (TMDs) in terms of polarized dipole amplitudes. These amplitudes extend the conventional small-$x$ eikonal dipole scattering framework by incorporating spin-dependent effects through energy-suppressed sub-eikonal and sub-sub-eikonal operators, embedded within newly defined polarized Wilson lines. Additionally, we have formulated small-$x$ evolution equations for each of these polarized dipole amplitudes. In the linearized large-$N_c$ limit, these equations become self-contained, enabling us to solve them by further imposing the double logarithmic approximation (DLA). The DLA equations can be implemented numerically, and in conjunction with the leading scaling behavior of the TMDs obtained here, are ready for application in various aspects of phenomenology. 

The set of leading-twist TMDs can be extracted from various spin-asymmetry observables in semi-inclusive deep inelastic scattering (SIDIS). Currently available data does not have significant coverage at small $x$, but the future Electron-Ion Collider (EIC) will probe down to $x\sim 10^{-3}$ \cite{Accardi:2012qut,Boer:2011fh,Proceedings:2020eah,AbdulKhalek:2021gbh,AbdulKhalek:2022hcn,Amoroso:2022eow,Abir:2023fpo,Burkert:2022hjz,Hentschinski:2022xnd}. So the small-$x$ evolution of TMDs will enhance the precision analysis of future EIC data, and open up several new observables for testing the polarized dipole picture and saturation physics. In addition, the evolution also provides input for global analysis, as has been demonstrated for quark and gluon helicity using explicit small-$x$ evolution in \cite{Adamiak:2023okq}. The asymptotic scaling results have also been applied as upper bounds on the growth of the transversity TMD at small-$x$ in \cite{Cocuzza:2023vqs,Cocuzza:2023oam}.

Aside from experimental probes of TMDs, lattice calculations can be used to compute moments of TMDs. As the moments involve an integration of the TMD down to $x = 0$, the lattice calculations of moments can serve as a cross-check against predictions from small-$x$ evolution. Lattice calculations of the TMDs themselves could provide initial conditions for small-$x$ evolution, allowing for full first-principles simulations of TMDs at moderate to low $x$.

We have numerically or analytically extracted the small-$x$ asymptotic scaling of six leading-twist flavor singlet quark TMDs and presented them together with the known results for the unpolarized and helicity TMDs in Table~\ref{tab_tmds}, from which we can see several interesting patterns emerge. Going left to right through Table~\ref{tab_tmds}, from unpolarized to longitudinally and then transversely polarized quark, one sees that the scaling generally becomes more and more energy suppressed similarly to the na\"ive scaling of the TMDs, with the exception being the slower growth of the unpolarized  TMD $f_1$.

Going column by column, we consider the unpolarized quark distributions, $f_1$ and $f_{1T}^{\perp}$. We make use of the fact that QCD is parity even and of CPT conservation. By looking at the operators from which we construct the TMDs, we can determine whether they are na\"ive T-odd or T-even distributions. Thus, one can infer the dominant charge parity of the TMD and this has implications for the relative growth of the singlet vs the non-singlet distributions. The unpolarized quark TMD, $f_1$, is a na\"ive T-even function, and its flavor singlet term comes from the eikonal dipole scattering amplitude, growing much faster than the flavor non-singlet contribution corresponding to the Reggeon. Conversely, the na\"ive T-odd Sivers function, $f_{1T}^{\perp}$, has a dominant eikonal contribution only in the C-odd flavor non-singlet sector. The flavor singlet contribution is actually sub-leading at small $x$, the exact opposite of the usual hierarchy found in small-$x$ QCD. This is because usual small-$x$ resummation involves C-even Pomeron exchanges, while the leading contributions to the Sivers function are C-odd and thus sensitive to the quark-antiquark imbalance encoded in the flavor non-singlet sector.

The second column consists of the quark helicity and transverse worm-gear TMDs. Here, we find that the flavor singlet contributions grow faster than the flavor non-singlet counterparts, as expected for T-even TMDs which should receive Pomeron like evolution corrections. Furthermore, the polarized dipole amplitudes and evolution equations are very similar between the two flavor singlet TMDs. 

Finally, in the third column, we have the Boer-Mulders function, the helicity worm-gear, the transversity and the pretzelosity. There are two notable patterns in this column, firstly that the small-$x$ asymptotic scaling for all four TMDs in this column is the same in both the flavor singlet and flavor non-singlet sectors. This can be traced back to the level of the polarized dipole amplitudes, where only polarized quark exchange operators contribute to the these TMDs at DLA in the massless quark limit. As a result, the evolution is restricted to take place via either eikonal emissions or polarized quark emissions. Such the restriction usually\footnote{Note that there can be exceptions for ``odderon-like" polarized gluon exchanges such as in the sub-eikonal contribution to the flavor non-singlet Sivers function \cite{Kovchegov:2022kyy}.} applies also for the evolutions in the flavor non-singlet sectors at large $N_c$. Physically, this feature is due to the fact that these TMDs are all chiral-odd and thus have no gluon TMD counterparts to mix with under the evolution. 

The second pattern we observe is that the TMDs which are ``off-diagonal" in spin -- the Boer-Mulders function, which encodes transversely polarized quarks in an unpolarized target, and the helicity worm-gear which encodes transversely polarized quarks in a longitudinally polarized hadron -- receive no $\mathcal{O}(\sqrt{\as})$ corrections to their na\"ive asymptotic scaling. Here, for the flavor singlet TMDs, we see that this only holds within the third column when the quark transverse polarization is probed. However, for the flavor non-singlet sector, this pattern extends to all of the ``off-diagonal" TMDs \cite{Santiago:2023rfl}. 

It is also interesting to note that the Sivers function and Worm-Gear $g_{1T}$ have multiple unique dipole amplitudes that are slow to converge to the same intercept. The dipole amplitude with the fastest growth should dominate the behavior of the TMD, but we find that this takes many decades in rapidity, well beyond the point that saturation corrections would become important and modify the power law growth of the dipole amplitudes.

Another important note is that bringing in contributions from finite quark mass, $m_f$, would allow for potential $m_f / k_T$-suppressed contributions, which are \emph{sub-eikonal} and therefore possibly parametrically comparable to the contributions considered here. In the case of the Boer-Mulders TMD, it was found that such corrections evolve trivially but enter at sub-eikonal order \cite{Kovchegov:2022kyy} as opposed to the sub-sub-eikonal contribution in the massless quark limit. This correction then comes parametrically as $m_f /k_T$ while the massless term scales as $x$, so the mass correction can become important depending on the kinematics of the process. We leave a detailed study of these quark-mass corrections for the other TMDs to future work. 

Achieving high-precision predictions necessitates relaxing several simplifying assumptions made in the derivation of the evolution equations. Progress in this direction has already been made for quark and gluon helicity TMDs. For example, the evolution equations have been generalized to the large-$N_c \& N_f$ Veneziano limit \cite{Cougoulic:2022gbk,Adamiak:2023okq}. In order to better estimate theoretical uncertainties, sub-leading single logarithmic corrections to the evolution have also been investigated \cite{Kovchegov:2021lvz}.

Efforts have also been directed at generalizing the equations to all $N_c$ values to formulate a Jalilian-Marian-Iancu-McLerran-Weigert-Leodinov-Kovner (JIMWLK)-type equation \cite{Jalilian-Marian:1997jx,Jalilian-Marian:1997gr,Jalilian-Marian:1997dw,Iancu:2001ad,Iancu:2000hn}. This involves incorporating sub-eikonal operators required for helicity-dependent insertions within the Wilson lines, as studied in \cite{Cougoulic:2019aja,Cougoulic:2020tbc}.

There is another important direction for future development. In this work, we have obtained evolution equations which resum the logarithms of $1/x$ which arise in TMDs, but there are also logarithms of the renormalization scale $\mu^2$ resummed by CSS evolution \cite{Collins:1981uw,Collins:1981uk,Collins:1981va,Collins:1984kg,Collins:1989gx}. In order to connect our small-$x$ formalism to the usual TMD factorization, we need to match the two sets of logarithms in the kinematic domain where both the collinear and eikonal approximations are valid. For scattering with both the hard scale $k_T^2 \sim Q^2$ and the center of mass energy $s$ sufficiently large, one must resum leading double logarithms $\alpha_s \ln (1/x) \ln (Q^2/\mu^2)$ in both the TMD factorization and small-$x$ formalisms. For the diagonal entries in \tab{tab_tmds} we can simply compare with the DGLAP evolution logarithms \cite{Gribov:1972ri,Altarelli:1977zs,Dokshitzer:1977sg} of the corresponding collinear PDF. This matching has been performed in both the unpolarized \cite{Ryskin:1980yz,Levin:1980za} and helicity TMD cases \cite{Adamiak:2023okq,Borden:2024bxa}. However, for the off-diagonal entries, the scale evolution which will be matched is that of twist-three collinear PDFs \cite{Braun:2009mi}. Here we have two connections to make: firstly we should construct the small-$x$ limit of the twist-three PDFs in terms of operators and match them to the corresponding TMD operators, checking against the relations which have been derived in the TMD factorization framework \cite{Moos:2020wvd,Rein:2022odl}. Second, having matched the operators we can compare the logarithms obtained from our small-$x$ evolution equations to those resummed by leading-order (LO) scale evolution of the twist-three PDFs. The first step will require the novel application of LCOT to twist-three PDF operators, possibly involving new polarized dipole amplitudes beyond what enters the corresponding leading-twist TMDs, and checking the higher order logarithms generated by small-$x$ evolution against collinear evolution would require NLO twist-three PDF evolution kernels, which have yet to be constructed. Such matching has been investigated recently for unpolarized scattering processes \cite{Fu:2024sba}, but spin-dependent scattering, including processes which would be sensitive to the distributions which match onto the off-diagonal entries in \tab{tab_tmds}, have yet to be addressed. Thus, we leave this considerable but crucial task of matching to future work. 

For full phenomenological applications and global analyses, it is also essential to compute the dipole amplitudes contributing to gluon TMDs at small-$x$ and derive their corresponding evolution equations. Addressing these challenges represents key areas for future research.


\section*{Acknowledgements}
We thank Yuri Kovchegov and Zhite Yu for useful discussions and reviewing drafts of this letter. We thank Ralf Seidl for discussions on EIC uncertainties and transversity measurements.

This work was supported by the Center for Nuclear Femtography, Southeastern Universities Research Association, Washington, D.C. and U.S. DOE Grant number DE-FG02-97ER41028 (MGS), and also by the U.S. Department of Energy, Office of Science, Office of Nuclear Physics under Award Number ~DE-AC05-06OR23177 (DA, MGS) under which Jefferson Science Associates, LLC, manages and operates Jefferson Lab.

YT was supported by the Research Council of Finland, the Centre of Excellence in Quark Matter and projects 338263, 346567 and 359902, and by the European Research Council (ERC, grant agreements No. ERC-2023-COG-101123801 GlueSatLight and No. ERC-2018-ADG-835105 YoctoLHC). The content of this article does not reflect the official opinion of the European Union and responsibility for the information and views expressed therein lies entirely with the authors. 


%

\end{document}